\documentclass[a4paper,fleqn,usenatbib]{mnras}

\usepackage[T1]{fontenc}
\usepackage{ae,aecompl}

\usepackage{graphicx}	
\usepackage{amsmath}	
\usepackage{amssymb}	
\usepackage{tikz}

\title[Spectral Study of S0s]{A SALT Spectral Study of S0s Hosting Pseudobulges}

\author[Vaghmare et al.]{
  Kaustubh Vaghmare$^{1}$ \thanks{E-mail: kaustubh@iucaa.in},
  Sudhanshu Barway$^{2}$ \thanks{E-mail: sudhanshu.barway@iiap.res.in},
  Petri V\"ais\"anen$^{3,4}$,
  \newauthor
  Rajin Ramphul$^{3,5}$,
  Yogesh Wadadekar$^{6}$, and
  Ajit K Kembhavi$^{1}$
  \\
  $^{1}$ Inter-University Centre for Astronomy \& Astrophysics
  (IUCAA), Post Bag 4, Pune 411007, India \\
$^{2}$ Indian Institute of Astrophysics, II Block Koramangala, Bangalore
560034, India \\
$^3$South African Astronomical Observatory, P.O.Box 9, Observatory 7935, Cape Town, South Africa,\\      
       $^4$Southern African Large Telescope, P.O.Box 9, Observatory 7935, Cape Town, South Africa,\\    
  $^{5}$ University of Cape Town, Astronomy Department, Private Bag X3, Rondebosch 7701, South Africa \\
  $^{6}$ National Centre for Radio Astrophysics (NCRA), Post Bag 3, Ganeshkhind, Pune 411007, India \\}

\begin{document}
\label{firstpage}
\pagerange{\pageref{firstpage}--\pageref{lastpage}}
\maketitle

\begin{abstract}
We present a SALT-RSS spectroscopic study of a sample of S0 galaxies
established by \citet{Vaghmare2015} as having pseudobulges using
a combination of photometric criteria. We extract the spectra of various regions
along the galaxy major axis  using standard long-slit spectroscopic reduction procedures and
model the spectra using STARLIGHT to derive detailed star formation histories. 
The central spectra of galaxies without bars in our sample reveal a complex star formation history, which is consistent 
with the belief that pseudobulges have a history of  star formation distributed over extended periods of time.  The
spectra of the unbarred galaxies  contain strong emission lines such as
$H\alpha$, indicating active star formation, which appears to be in contradiction with the expectation that 
S0  galaxies have been stripped of gas.  In the case of the two barred galaxies
in the sample,
the spectrum is dominated by light
from a much older stellar population. This seems to suggest an accelerated
formation of the pseudobulge made possible by the action of the bar. One of
these galaxies appears to have exhausted its reservoir of gas and thus has no
signature of a recently formed population of stars while the other galaxy has
managed to give rise to new stars through a recent funnelling action.
We have also confirmed the influence of bars on the nature of the stellar
population in a pseudobulge using an alternate sample based on the SDSS.

\end{abstract}

\begin{keywords}
methods: data analysis - techniques: image processing - catalogues - galaxies: general.
\end{keywords}



\section{Introduction}
\label{sec:intro}

There are several ways by which one can approach the study of galaxy
evolution. One approach is to classify the galaxies in the nearby
Universe into different morphological types and study their
properties. In such an approach, the study of S0 galaxies has received
considerable attention in the last decade. S0 galaxies are
characterised by the presence of, what appears to be, an
elliptical-like bulge and a smooth outer disk devoid of any apparent
spiral arms. Indeed, the study of these galaxies is important because
hidden within their inter-relationship with galaxies of other types
are clues to the varied processes that play important roles in the
overall picture of galaxy evolution (See \citet{Aguerri2012} and
references therein).

For example, observational studies suggest that the central components
of S0 galaxies viz. bulges resemble elliptical galaxies
\citep{Renzini99}.  They have thus likely formed in a similar manner
to ellipticals, i.e. through hierarchical clustering and
mergers. However, the observation that the fraction of S0 galaxies
within a cluster of galaxies grows at the cost of spirals
\citep{Dressler1980}, suggests that S0s originate from spirals. Gas
stripping via ram pressure is often considered a viable process for
such a transformation in dense cluster environments \citep{Gunn1972,
Aragon-Salamanca2006, Laurikainen2010, Maltby2015}. However, we know
that S0's are equally common in groups \citep{Wilman2009, Just2010,
  Bekki2011}. These observations suggest that the processes such as
ram pressure stripping are not responsible for the formation of the
majority of S0s in groups where minor mergers and tidal effects can
control their evolution \citep{Mazzei2014, Mapelli2015}.

Simulations too suggest that there are multiple methods for forming
S0s - they can arise in major mergers, can grow through slow/secular
processes and can also form through gas stripping of spirals
\citep{Querejeta2015, Tapia2017}. Traces of past merger events which
impact the observed kinematics have been observed in some S0s
\citep{Falcon-Barroso2004}. The observed star formation (SF) and SF 
history (SFH) characteristics of S0s can be complex 
(e.g. \citet{Barway2013} and references therein).

\citet{Barway2007, Barway2009} suggest that the dominant formation
process  in case of S0s is a function of luminosity, with brighter S0s 
forming in a manner similar to ellipticals and fainter S0s forming
through secular processes or perhaps by gas stripping of spirals
\citep{Rizzo2018},
which in turn may have formed through secular evolution. 
\citet{Vaghmare2013} used archival deep mid-infrared images from the 
Spitzer Infrared Array Camera to systematically study the bulges of S0 
galaxies. Bulges, as defined photometrically, are excesses of light
over  an inward extrapolation of an exponential profile fitted to the 
outer disk. They are known to occur in two varieties - classical,
resembling ellipticals in almost every aspect including morphology, 
colour, nature of stellar populations and kinematics,  and pseudo, 
resembling disks and believed to have formed through secular 
evolution \citep{Kormendy2004}. \citet{Vaghmare2013} point out
that pseudobulges do exist in S0 galaxies and preferentially
occur in fainter S0s. But what is perhaps a more interesting result is
that disks of pseudobulge hosting S0s have a lower scale length than
those hosting classical bulges.

In \citet{Vaghmare2013}, the authors speculate that the lower scale length could imply either
(i) that the population of disks with lower scale length on average preferentially
host pseudobulges or (ii) that the lowered scale length is a result of some process
responsible for either the bulge's or the overall galaxy's growth. In a
follow-up study, \citet{Vaghmare2015} compare the disk properties of spirals and
S0s and find that pseudobulge hosting S0s have a lower scale length than the
pseudobulge hosting spirals. The authors explain this as being due to lowered disk
luminosity in case of S0s hosting pseudobulges. They further compare bulge
luminosities to demonstrate that only the early-type spirals may transform into
S0s with pseudobulges through disk fading via gas stripping, while late-type
spirals will need additional processes, especially to aid the growth of bulges to
observed luminosities.

The conclusions reached by \citet{Vaghmare2013, Vaghmare2015} and by Barway et
al. rely on statistical arguments. Correlations are derived between
various parameters, which in turn are derived from imaging data. While such
studies indeed offer insights into broad trends among galaxies in a given class,
they cannot provide information on individual galaxies and their star formation /
assembly histories. For this, one requires data from multiple wavebands. 
\citet{Barway2013} use broad band photometry from multiple surveys
spanning UV, optical and infrared wavelengths but again use statistical
arguments to point out broad trends among S0 galaxies. In order to truly
constrain the formation histories of S0s, one needs to have spectroscopic data.
Using stellar population synthesis techniques, one can then derive very detailed
star formation histories for these galaxies and confront the conclusions reached
by the photometric studies. To this effect, we have been using the Southern
African Large Telescope's (SALT) Robert Stoby Spectrograph (RSS) to observe a
select sample of S0 galaxies.

The work presented in this paper
focusses on a subset of the galaxies - those shown to host pseudobulges by
\citet{Vaghmare2015}, whose photometric properties and their implications were
summarised earlier. The aim here is to check
for consistency between star formation histories and the conclusions based on
the photometric observations. This paper is organised
as follows - the data and the reduction methods are described in section
\ref{sec:data}, section \ref{sec:specmod} explains the process of modelling the
spectra using \textsc{starlight} while sections \ref{sec:result}, \ref{sec:discussion}
and \ref{sec:bar_dn4000}
present the detailed results and possible interpretations.

Throughout this paper, we use the standard concordance cosmology with $\Omega_M=
    0.3$, $\Omega_\Lambda= 0.7$ and $h_{100}=0.7$. Also, magnitudes used
in this work are in the AB system.

\begin{figure*}
\centering
\includegraphics[width=0.8\textwidth]{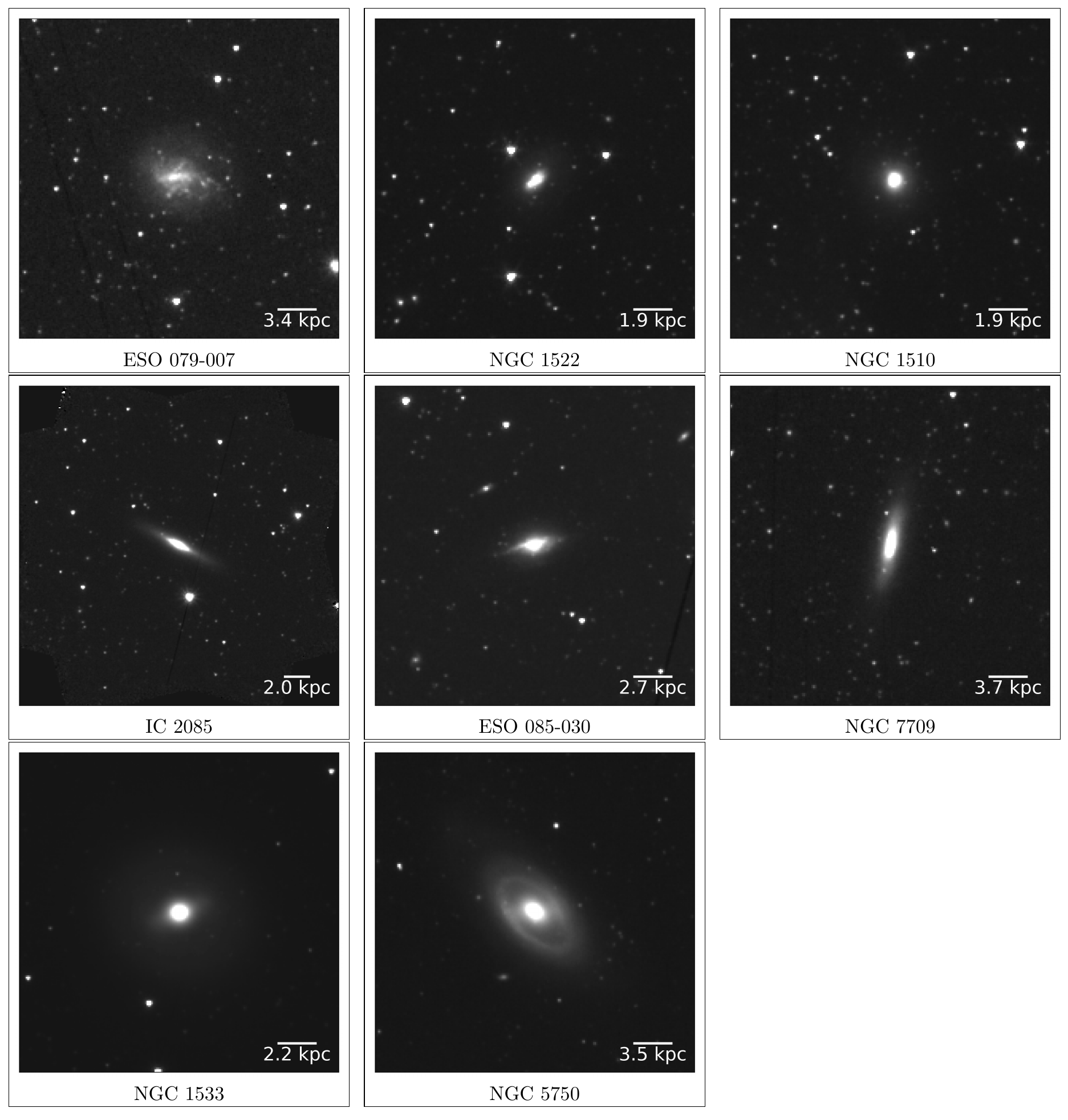}
\caption{The 3.6 micron images obtained using the Spitzer IRAC, for the eight
    galaxies studied in the present paper.}
    \label{fig:spitzer_images}
\end{figure*}

\section{Sample and Observations}
\label{sec:data}

\begin{table*}
\begin{center}
\begin{minipage}{166mm}
\caption{Basic parameters of the sample galaxies.}
\label{tab:pb_observations}
 \begin{tabular}{l c c c c l}
 \hline
Object Name &  RA  (J2000) & Dec (J2000) & Total magnitude  & Exp. Time & Position Angle \\
 &  hh:mm:ss & dd:mm:ss & B band & seconds & degree \\
\hline
ESO079-007  & 00:50:04 &  -66:33:10 &  13.73  &  4500 & 2.6 \\
NGC1522       & 04:06:08 &  -52:40:06 &  13.97  &  5000 & 38.6 \\
NGC1533       & 04:09:52 &  -56:07:06 &  11.79  &  3600 & 170 \\
NGC1510       & 04:03:33 &  -43:24:01 &  13.47  &  4500 & 102.9 \\
IC2085           & 04:31:24 &  -54:25:01 & 13.95   &  5000 & 112 \\
ESO085-030  & 05:01:30 &  -63:17:36 &  13.79  &  4500 & 148.2 \\
NGC5750       & 14:46:11 &  -00:13:23 &  12.55  &  4500 & 70.6 \\
NGC7709       & 23:35:27 &  -16:42:18 &  13.60  &  4500 & 55 \\
\hline
 \end{tabular}

\textbf{Notes:} Column 1 is the galaxy name ; Column 2 \& 3 are Right
 Ascension and Declination of galaxy ; Column 4 is the total B band
 magnitude; Column 5 is the total exposure time for RSS spectrum and
 Column 6 is the slit position angle, measured North to East.
\end{minipage}
\end{center}
\end{table*}

The sample used in the present study is described in detail in
\citet{Vaghmare2013} and \citet{Vaghmare2015}. For convenience, 
we summarise its essential features here. The parent sample comprises 
3657 S0 galaxies as found in the \emph{Third Revised Catalogue}
\citep{deVauc91}. A total B-band magnitude cut-off of 14.0 was imposed 
to reduce this sample to 1031 galaxies which were then cross-matched
with the Spitzer Heritage Archive (SHA). We found mid-infrared images
taken by the Spitzer Infrared Array Camera (IRAC) for 247 galaxies at 
3.5 $\mu m$. These images were processed to achieve the desired
quality for 2D decompositions and structural parameters were obtained 
by using GALFIT \citep{Peng2010} for a subset of 185 S0s. Of these, 25 
were classified as having pseudobulges using a very conservative 
criterion -- bulges should lie more than 3-$\sigma$ below the Kormendy 
relation  \citet{Kormendy2004} and should have an S\'{e}rsic index
$n<2$.  This is a combination of the criteria used by
\citet{Gadotti2009} and \citet{Fisher2008} (see
\citet{Vaghmare2013} for full details).

We used the 10m class Southern African Large Telescope (SALT) \citep{Buckley2006} and its
Robert Stobie Spectrograph (RSS)\citep{Burgh2003, Kobulnicky2003}  to study the sample. The RSS is
a versatile instrument providing several observing modes and is designed to work in a wavelength range of 320 - 900 nm.
Of the 25 pseudobulges from \citet{Vaghmare2015}, 15 were in the
declination range accessible to SALT.  We were given time to observe 8
of these galaxies which we present in this paper. A brief summary of the
observational parameters is presented in Table \ref{tab:pb_observations}. 
Spitzer 3.6 micron imaging for the  galaxies is shown in Figure  \ref{fig:spitzer_images}.

The long-slit RSS spectra were obtained using a 1'' wide slit and the PG0900 grating, placing the slit 
along the major axis of the galaxy in each case.
RSS uses an array of three
CCDs of $2048 \times 4096$ pixels, and the grating angle in the instrument  was set in a way 
so as to avoid any important spectral features falling in the resulting two 15 pixel size gaps.
We used a $2 \times 4$ CCD binning scheme to improve the
signal-to-noise ratio (SNR). The spectra obtained in the said configuration have an observed 
wavelength range of 3640-6765 \AA  \ and thus cover the required optical range 
for unravelling the stellar formation histories of these galaxies. The spectral resolution is $\sim$ 3 \AA. 

We used multiple 600-900 sec individual exposures for achieving the total
exposure times per galaxy as listed in Table~\ref{tab:pb_observations}.  
For 7  of the  8  galaxies in our sample, this was done over two 
separate visits to the targets.  Calibration unit Arc lamps and flat-field images were observed with the same configuration as
used for the science targets and spectrophotometric standards.

\subsection{Spectroscopic Reduction}

We made use of the SALT \emph{product data} generated by the in-house pipeline called
PySALT \citep{PySALT}, which mosaics the individual CCD data to a single FITS file, corrects for cross-talk effects, 
and performs bias and gain corrections. We then carried out further reduction
steps using our own custom tools,
written in Python/PyRAF, which consist of several
modules designed to handle specific steps of processing.  
First, we trim the regions of the CCD not containing any usable
data and then fill the CCD gaps using a gradient fill. To do this, we
    calculate the
    the gradient across the gap by measuring the flux values of
    the pixels to the right and the left of the gap, computing the difference
    and normalising it to the number of pixels belonging to the CCD gap (the
    gradient). We then use the gradient to determine the flux values for
    pixels in each row of the image and replace the zero value pixels by
these flux values. This step is
important to prevent discontinuities in the brightness distribution which 
affect the flat-fielding process which relies on the modelling of the large
scale illumination structure. Another advantage of this gradient fill is
    seen in the subsequent step of background subtraction. A CCD gap which
    appears straight in the original images becomes curved as a result of the
    coordinate transformation step. This causes background subtraction to result
    in discontinuous patches. By filling the CCD gap, these patches can be
avoided.
We then remove the cosmic rays using the LACosmic
algorithm \citep{LACosmic}. Calibration in wavelength is achieved using arc
lamps and has typical error of $\pm 0.35$ \AA. We use the arc lamp to
determine a coordinate transformation to `straighten' frames so that every 
column corresponds to a unique wavelength making the background sky shape 
fitting and subtraction possible.   Using the spectrophotometric
standard data, we determine the relative flux calibration. We then check
individual frames for alignment and co-add them. During this step, 
a standard deviation frame is constructed from which an error frame is obtained. 
Before the co-addition, to have consistent noise characteristics, we equalize
the effective exposures of the frames --  this is needed because the pupil of SALT
changes by design during the observation. 
If $O_i(x, y)$ are
the individual observed frames, the final spectrum $F(x,y)$ and error spectrum
$E(x, y)$ can be written as, 

\begin{equation}
    F(x,y) = \left< O(x,y) \right> = \frac{\sum_i^N O_i(x, y)}{N}
    \label{eq:final_frame}
\end{equation}

\begin{equation}
    E(x,y) = \sqrt{\frac{\sum_i^N (O_i(x,y) - \left<O(x,y)\right>)^2}{N-1}}
    \label{eq:error_frame}
\end{equation}

An alternate way of obtaining the error frames is to start from the raw CCD
images whose noise properties are relatively well understood to try to model the effects of
every single reduction step on the errors. This is however very complicated and
so
we adopt the
above technique of determining our final error frame. Since we are combining
only 6 - 8 frames, we might overestimate our errors, but this
over-estimation does not affect the findings presented in the
paper. Our final step is to correct for foreground extinction and this is done using
the \emph{deredden} task in IRAF.

\subsection{Post Reduction Procedure}

\begin{figure}
  \centering
  \includegraphics[width=0.4\textwidth]{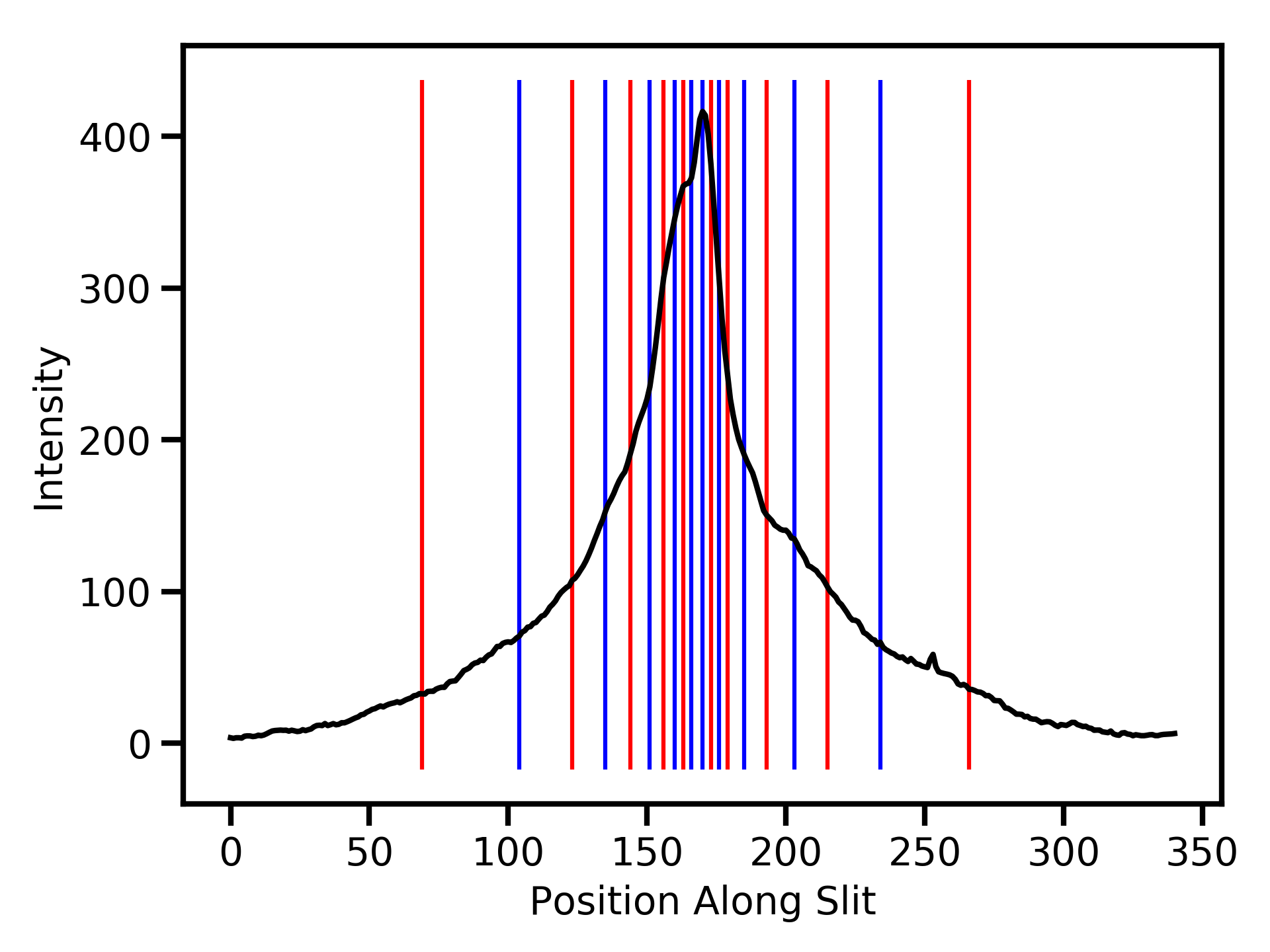}
  \includegraphics[width=0.4\textwidth]{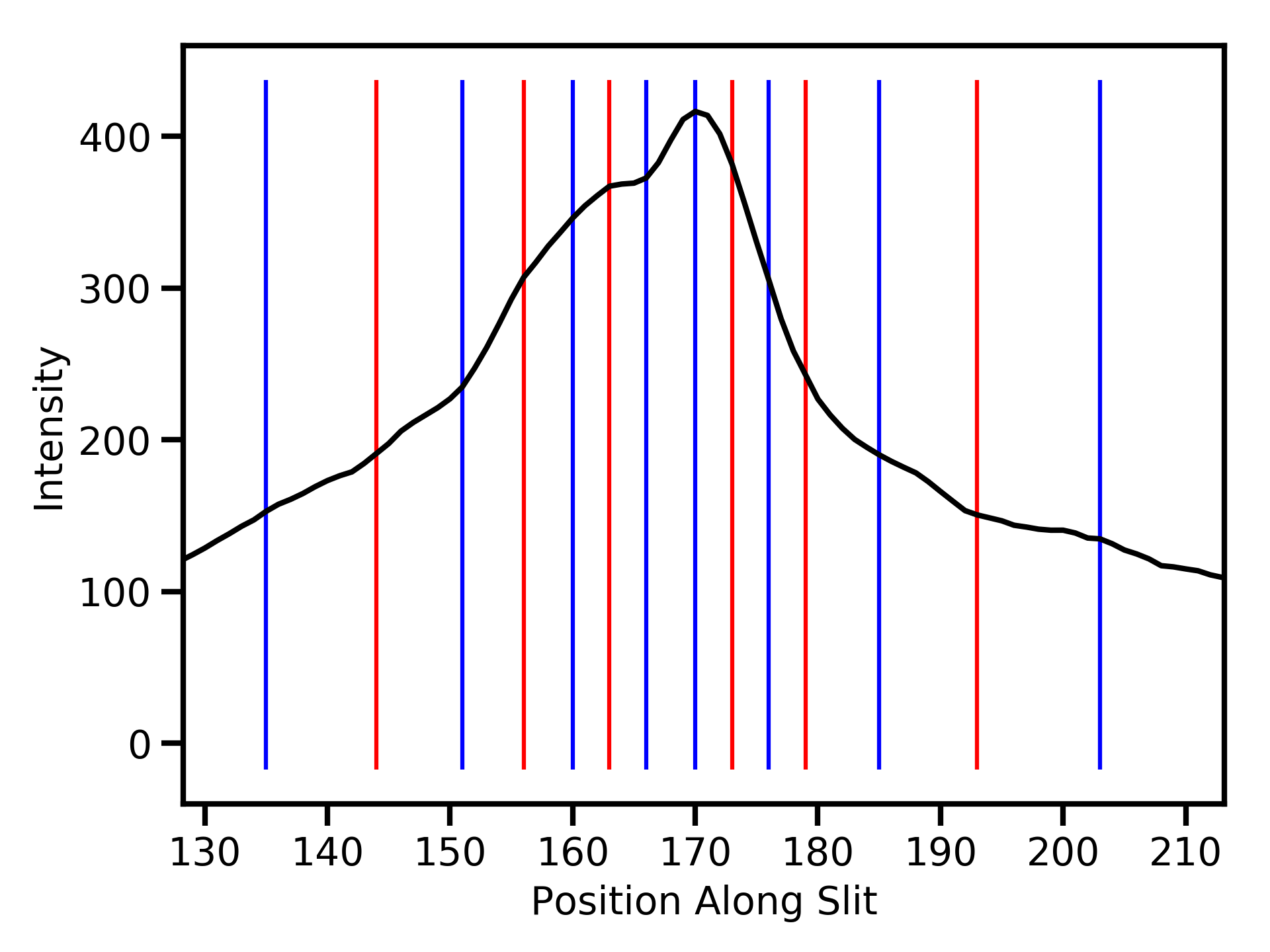}
  \caption{A plot of the brightness profile of a galaxy along the slit. The
  apertures, rows which are summed up to produce the 1-d spectrum in the
  given region of the galaxy, are marked in alternating colors. The rows lying
  between a pair of successive red and blue lines are summed to obtain the
  spectrum corresponding to the centre of this aperture.
  As can be seen, the width of these apertures increases in inverse ratio to the level of flux.
  The lower panel shows a close-up of the central region.}
  \label{fig:aperture}
\end{figure}
 
The modelling of the spectrum to study its star formation history has been
done using \textsc{starlight} \citep{Fernandes2005}. This program requires a 1-d spectrum
in the form of an ASCII table comprising at least two columns, namely wavelength
and flux.  Furthermore, we wish to study the spectra at different spatial positions. 
We thus extract 1-d spectra from positions along the
long slit by summing up a suitable number of rows. If this number
is kept constant as a function of the position on the long slit, the 
signal-to-noise ratio degrades for spectra extracted far away from the centre due to
the decreased brightness of the galaxy in this region. So, we came up with a
scheme of increasing the number of rows summed up as an inverse function of the
total amount of flux from the galaxy in a given region of the slit. This is
illustrated in Figure \ref{fig:aperture}, which shows the flux in scaled
counts as a function of position along the slit. The routine that
determines the sizes of these apertures requires as an input the number of
apertures to be extracted and the range of the spatial intensity profile over which the
extraction is to be carried out, and then determines these aperture sizes in units
of pixels so as to keep the total flux within the aperture roughly constant.
The routine does not take into account the physical scale of the galaxy. The
number of apertures to be extracted are chosen subjectively based on the overall
quality of the spectrum which in turn depends on the brightness of the galaxy.

Due to the rotation of stars in the galaxies, the spectra
represented by individual rows are shifted with respect to each other. 
We constructed a Python based module to characterise this by tracing
the centroid of individual absorption features along the long slit.  A rotation curve 
of the galaxy is thus obtained. By fitting polynomials of suitable order to this
curve, it is also possible to \emph{derotate} individual rows for the purpose of
stellar population fitting -- this is necessary to not introduce broadening
of spectral features due to rotation. We shift each row in the wavelength space
using the fitted rotation curve information before co-adding rows to be fit with \textsc{starlight}.
During the coaddition of the rows, the corresponding rows from the error frame are also
added in quadrature and divided by number of rows to get the error spectrum. 
The spectrum and the related uncertainties can now be read and modelled
by \textsc{starlight}.

\section{Spectral Modelling}
\label{sec:specmod}

There are several methods to derive the star formation history of a galaxy
from its spectrum (see \citet{Sanchez2016} for a critical summary). 
In the present study, we adopt the method implemented in \textsc{starlight}
\citep{Fernandes2005}. We briefly explain the method for the convenience
of the reader.

Any arbitrary star formation history can be approximated as a series of star
bursts in each of which a population of stars with a given initial mass function (IMF)
is formed almost at the same time. The spectrum of such a population, known as
a Simple Stellar Population (SSP), can be determined as a function of age and
metallicity, knowing the IMF and adopting a stellar spectral library. The
observed spectrum can then be modelled as a linear combination of spectra of
different SSP. A galaxy whose formation is consistent with a single
episodic burst of star formation would be described by a single SSP and thus any
attempt to model such a galaxy spectrum would lead us to finding virtually no 
contribution from any SSP but one. And for any other formation history, the
relative contributions would be proportional to the strength of the star bursts
at each epoch. This way, the star formation history of the galaxies can be
obtained.

Such a method is implemented by \textsc{starlight} which accepts an input 1-d spectrum, a
set of base spectra and data flags/masks and fits a \emph{population vector} to
the observed spectrum, which tells us the relative contribution of each spectrum
in the base to the observed spectrum. Mathematically, the model spectrum can be
written as

\begin{equation}
  M_{\lambda} = \sum_{j=1}^{N_*} P_j L_{\lambda,j}^0 \ast G(v_*, \sigma_*)
  10^{-0.4A_{\lambda,j}} 
  \label{eq:starlight_model}
\end{equation}

\noindent Here, $P_j$ refers to fractional contribution of jth base spectrum to
the total spectrum. $L_{\lambda,j}^0$ refers to the jth base spectrum without any
extinction or kinematic filters applied, G refers to a Gaussian characterized by
a centroid $v_*$ and a width $\sigma_*$. The operation $\ast$ implies a
convolution. The last multiplicative factor is to account for intrinsic
extinction along the line of slight. $N_*$ refers to the total number of base
spectra.

However, an important issue concerning the
use of \textsc{starlight} is with regards to the base selection. A hand picked base may be biased
towards certain ages and metallicities and thus introduce biases in the deduced
star formation history. An attempt to conservatively choose all possible ages
and metallicities increases computation overheads (as $N_*^2$). We thus adopt the prescription of \citet{Richards2009}
who suggest using diffusion mapping with K-means clustering to obtain a base of
M spectra starting from a very large number N of base spectra covering all or most of
the parameter space of interest, with $M \ll N$. For the present study we obtain
150 base spectra from the MILES \citep{MILES} spectral library obtained using the
Salpeter IMF,
 and use diffusion mapping with K-means clustering to construct
a base of 45 spectra, which uniformly covers the age-metallicity parameter
space. The initial base spectra are obtained from a web application provided by
the MILES group. The code for applying the \citet{Richards2009} algorithm is
provided by them as a MATLAB code. We wrote a converter to reformat the spectra
obtained from the MILES site into a form compatible with the MATLAB code. We
wrote another script in Python to transform the output of the MATLAB code into a
form compatible with \textsc{starlight}. The final 45 spectra used for modeling
the spectra of the galaxies is presented in Appendix A.

In order to compute the errors on the parameters returned by \textsc{starlight}, we adopt
the following approach. We use the derived error spectrum to construct several
($\sim 100$) realisations of the observed spectrum and run \textsc{starlight} for each
one of them. These realisations are constructed by approximating the
    underlying error distribution to be a Gaussian and drawing random samples
    from a Gaussian distribution with zero mean and a standard deviation equal
the corresponding uncertainty read off from the error spectrum. The standard deviation of any given parameter obtained for these
realisations is adopted as an estimate of its uncertainty. As noted earlier, the
method by which error frames are being constructed is likely to lead to an
over-estimation of the uncertainty. This means that the error bars obtained on
the \textsc{starlight} parameters will also be over-estimated. As can be gauged from the
discussion of the science results in the subsequent sections, this 
over-estimation does not affect the conclusions drawn in this paper.

\section{Results}
\label{sec:result}

\begin{figure*}
  \centering
  \includegraphics[width=0.9\textwidth]{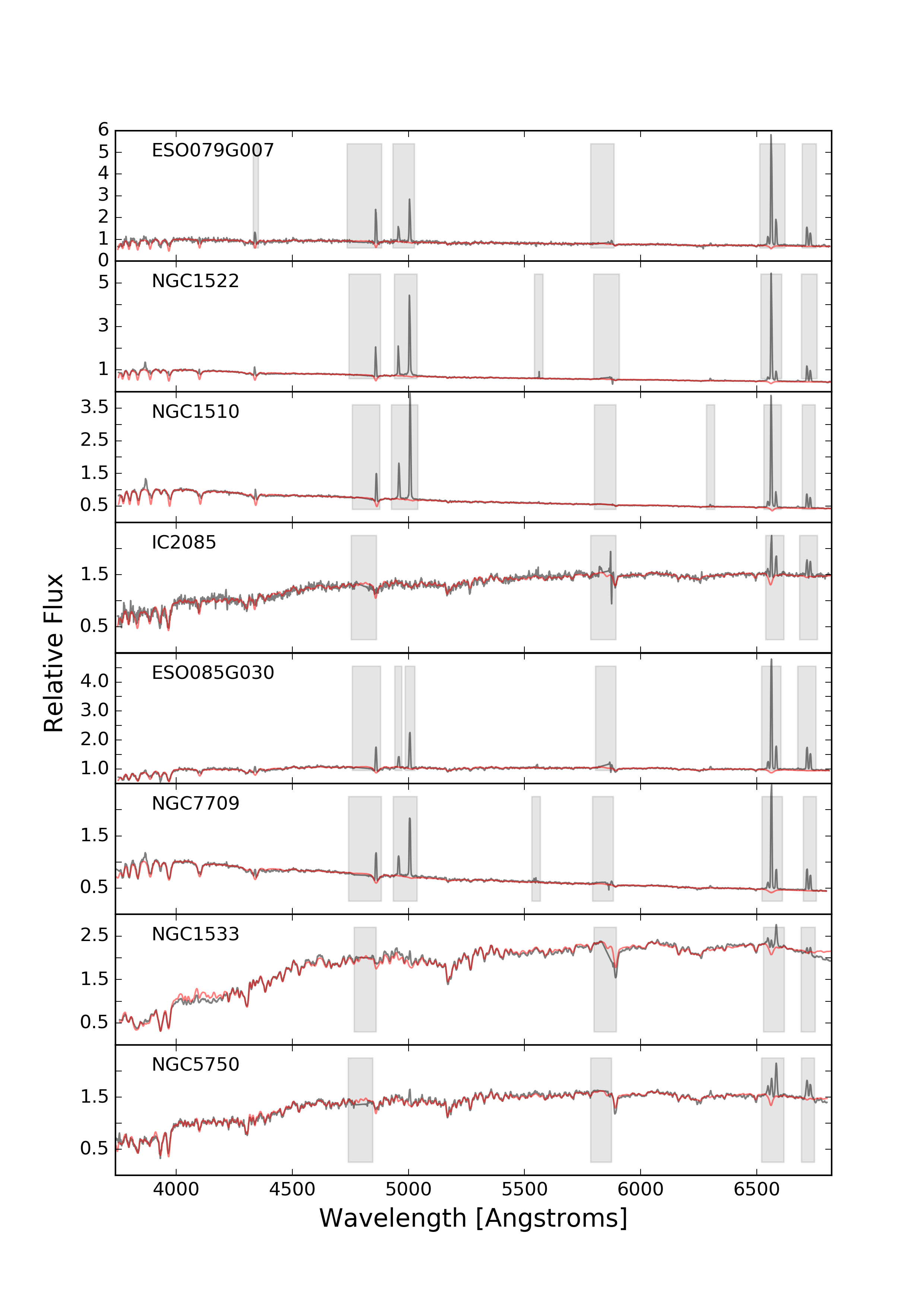}
    \caption{Observed spectrum (grey line) and best-fit \textsc{starlight} model
        (red line) for the central regions of galaxies in our sample.
    The grey regions indicate wavelengths ignored during the \textsc{starlight} fit as
    they contain emission from hot gas or are lost in the CCD gaps.}
  \label{fig:spectra_all}
\end{figure*}

\begin{figure*}
  \centering
  \includegraphics[width=0.9\textwidth]{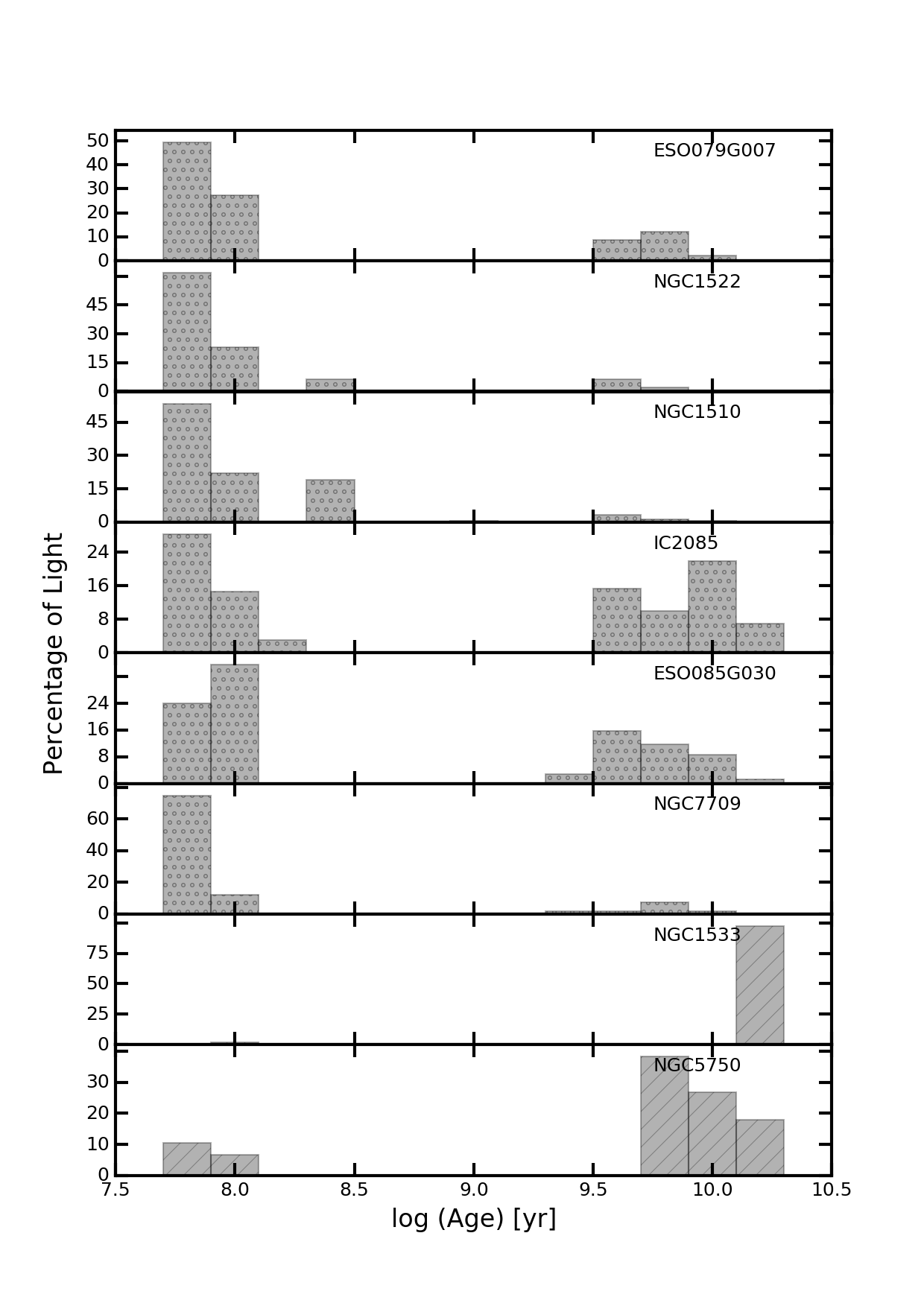}
  \caption{Star formation histories for the galaxies in our sample.
    The X-axis indicates the logarithm of the age and the Y-axis the
    percentage of light contributed to the observed spectrum by a stellar
    population of the given age. These percentages are computed from the
    population vector returned by \textsc{starlight}.The hatch patterns are circles for the first six
galaxies which are unbarred and forward leaning diagonals for the last two
galaxies which are barred.}
  \label{fig:sfh_all}
\end{figure*}

\begin{figure*}
  \centering
  \includegraphics[width=0.9\textwidth]{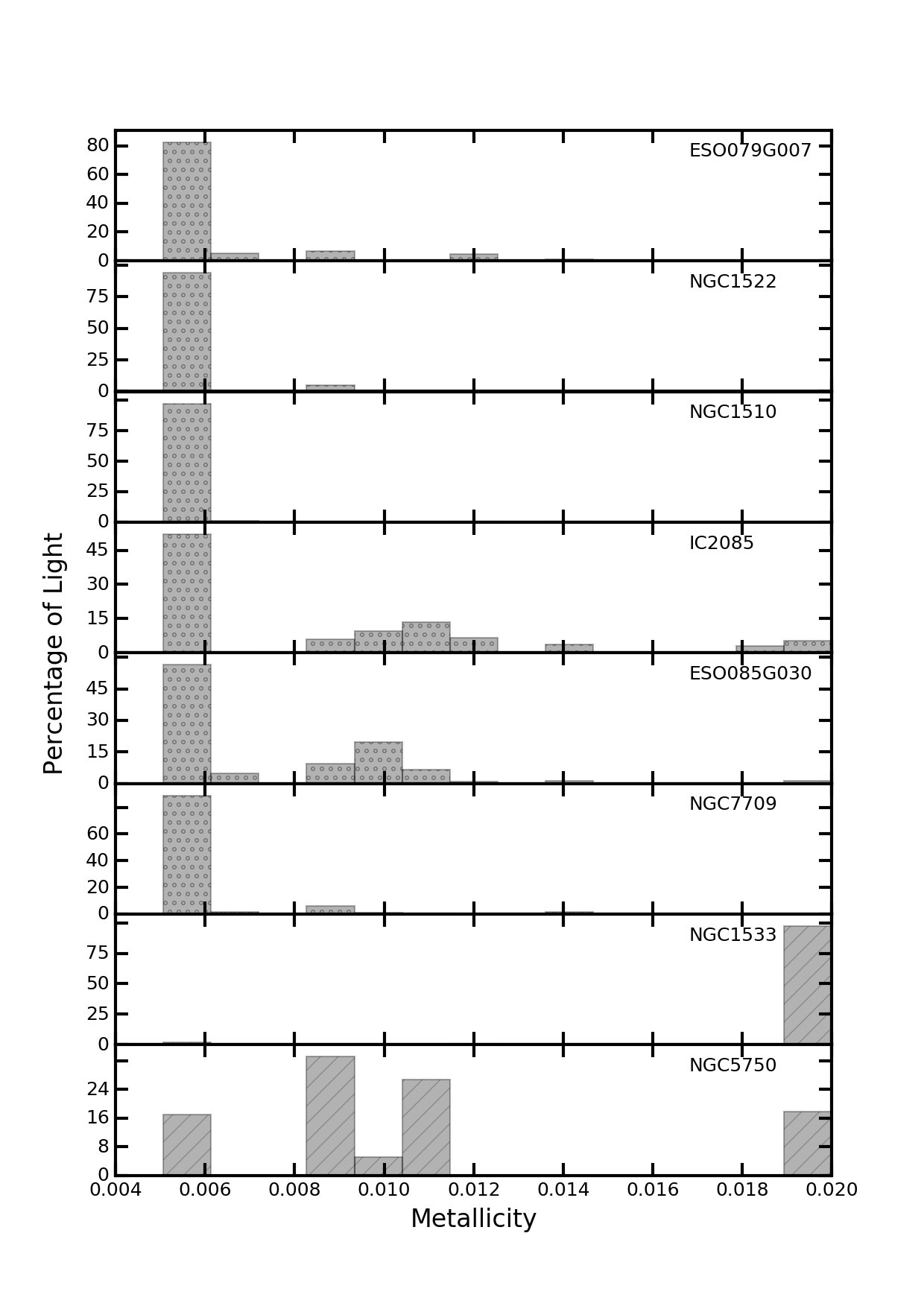}
  \caption{Metallicity contributions for the galaxies in our
    sample. The X-axis is the metallicity while the Y-axis represents the
    fractional contribution to the flux of the  observed spectrum as determined from the
population vector. The hatch patterns are circles for the first six
galaxies which are unbarred and forward leaning diagonals for the last two
galaxies which are barred.}
  \label{fig:metal_all}
\end{figure*}

\begin{figure*}
    \includegraphics[width=\textwidth]{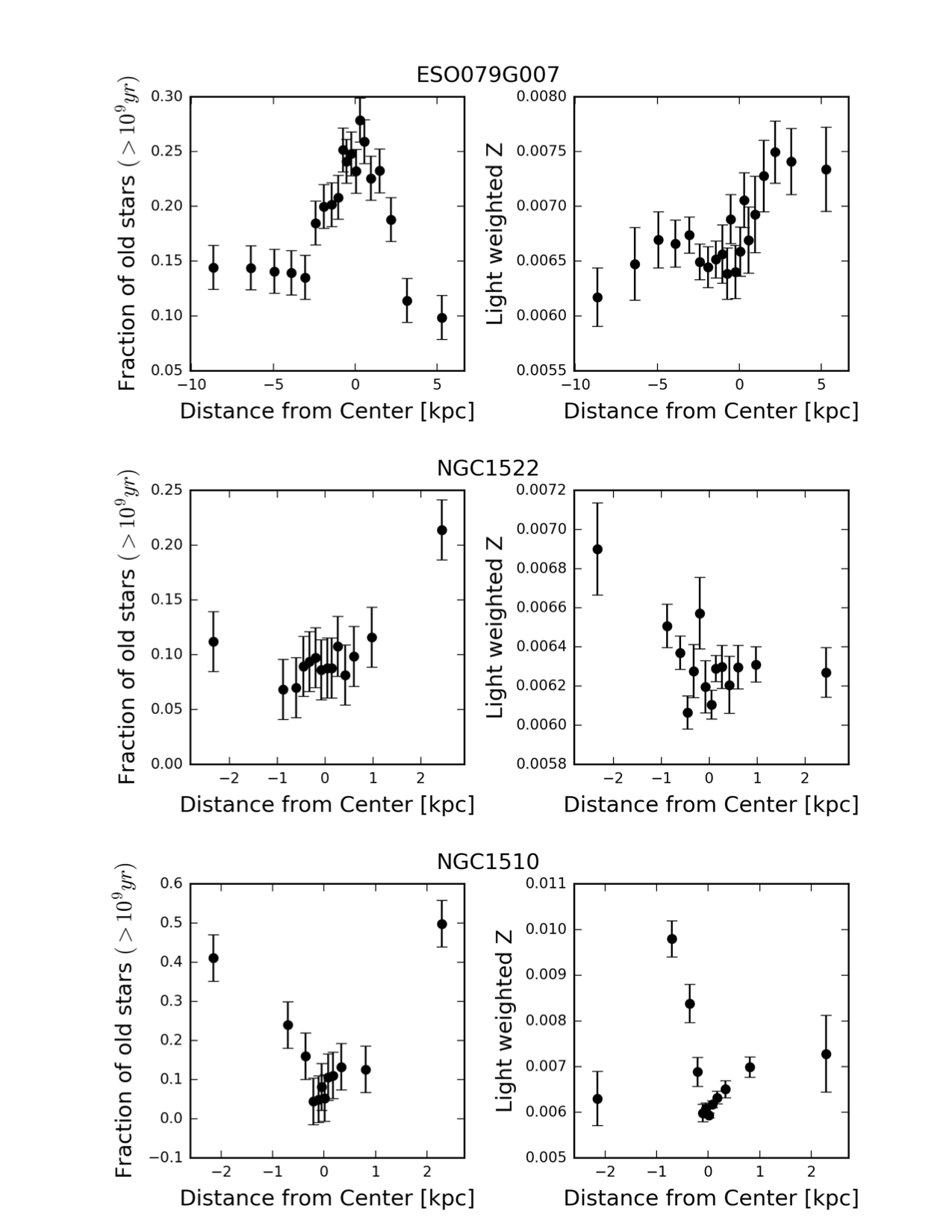} 
    \caption{Age and metallicity gradients for the galaxies in our sample. The
    X-axis in all the plots is the distance relative to the centre of the galaxy
in kpc. For plots on the left, the Y-axis indicates the fraction of the
stars with age $ > 10^9$ yr, while for the plots on the right, it indicates the light
    weighted metallicity. Both quantities are computed by \textsc{starlight} and can be
    obtained from the population vector output by it.}
  \label{fig:grads_others1}
\end{figure*}

\begin{figure*}
    \centering
    \includegraphics[width=\textwidth]{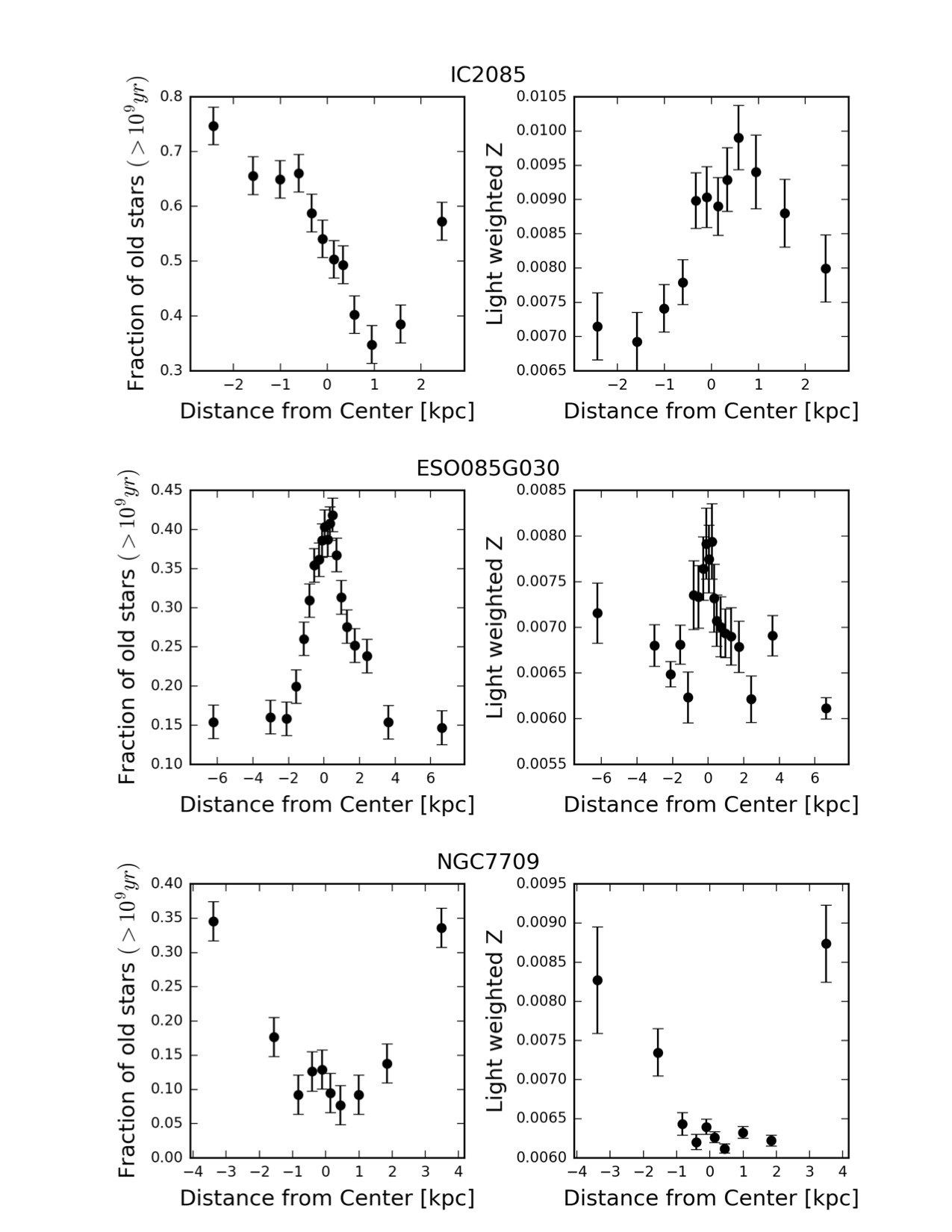} 
    \contcaption{Age and metallicity gradients for the galaxies in our sample. The
    X-axis in all the plots is the distance relative to the centre of the galaxy
in kpc. For plots on the left, the Y-axis indicates the fraction of the
stars with age $ > 10^9$ yr, while for the plots on the right, it indicates the light
    weighted metallicity. Both quantities are computed by \textsc{starlight} and can be
    obtained from the population vector output by it.}

  \label{fig:grads_others2}
\end{figure*}

\begin{figure*}
    \centering
    \includegraphics[width=\textwidth]{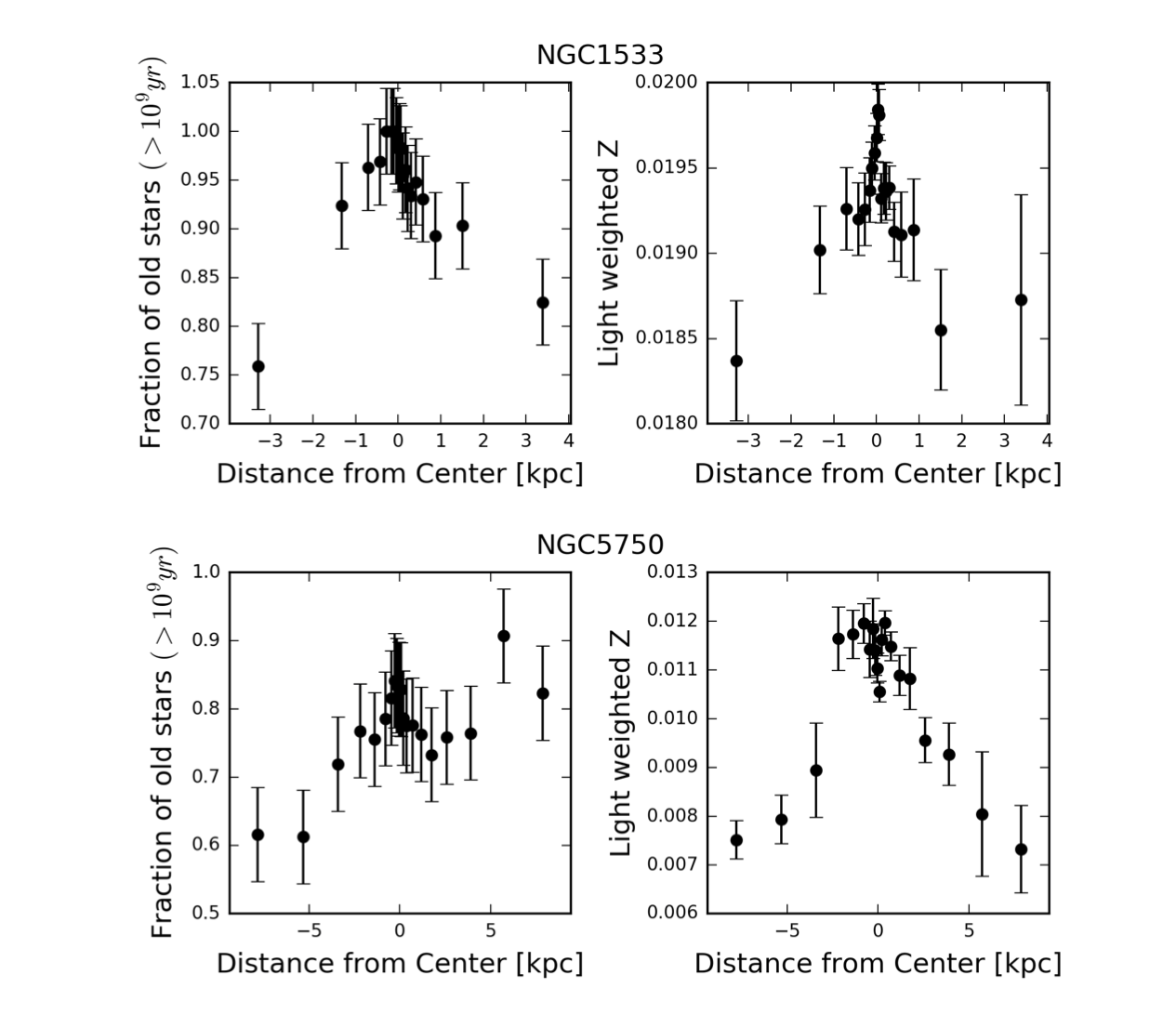} 
    \contcaption{Age and metallicity gradients for the galaxies in our sample. The
    X-axis in all the plots is the distance relative to the centre of the galaxy
in kpc. For plots on the left, the Y-axis indicates the fraction of the
stars with age $ > 10^9$ yr, while for the plots on the right, it indicates the light
    weighted metallicity. Both quantities are computed by \textsc{starlight} and can be
    obtained from the population vector output by it.}

  \label{fig:grads_barred}
\end{figure*}

\begin{table*}
\begin{center}
\begin{minipage}{166mm}
\caption{Best fit Bulge and Disk parameters for the S0 galaxies in our
sample.}
\label{tab:phot_params}
\begin{tabular}{l c c c c c c c c c c c c}
\hline
Name & $T$ & $L_K$ & \multicolumn{3}{c}{Bulge parameters} &
\multicolumn{2}{c}{Disk parameters} & \multicolumn{3}{c}{Bar Parameters}  & B/T
& Bar/T \\
 & & & $<\mu_e>$ & $\rm{r}_e$ & $n$ & $\mu_{0d}$ & $\rm{r}_d$  &
$\rm{m}_{bar}$ & $n_{\rm{bar}}$ & $\rm{r}_{e(\rm{bar})}$ & & \\
\hline
\hline
ESO079-007 & -2.0 & -18.73 & 23.25 & 0.87 & 0.53 & 21.23 & 1.72 & -- &
--
& -- & 0.04 & -- \\ 

NGC1522 & -2.3 & -18.33 & 20.63 & 0.36 & 0.44 & 21.83 & 0.81 & -- & --
&
-- & 0.37 & -- \\ 

NGC1510 & -1.8 & -20.24 & 19.18 & 0.24 & 1.18 & 21.41 & 0.82 & -- & --
&
-- & 0.40 & -- \\ 

IC2085 & -1.2  & -18.48 & 20.37 & 0.53 & 0.88 & 21.79 & 1.24 & -- & -- &
-- & 0.41 & -- \\ 

ESO085-030 & -0.4 & -18.76 & 20.50 & 0.86 & 1.41 & 20.94 & 0.78 & -- &
-- & -- & 0.64 & -- \\ 

NGC7709 & -1.9 & -20.92 & 20.65 & 0.72 & 0.41 & 21.03 & 1.70 & -- & --
&
-- & 0.20 & -- \\ 

NGC1533 & 0.1 & -18.79 & 16.73 & 0.22 & 0.95 & 19.77 & 1.58 & 12.94 & 0.50
&
0.89 & 0.22 & 0.07 \\ 

NGC5750 & -2.0 & -21.62 & 17.84 & 0.32 & 0.98 & 19.94 & 2.56 & 13.64 & 0.20
& 2.23 & 0.09 & 0.09 \\ 
 
 \hline
\end{tabular}

\textbf{Notes:} Column 1 - The common name of the galaxy; Column 2 - Hubble
stage parameter $T$, Column 3 - K-band absolute magnitude (AB system), Column 4 - the
average surface brightness of the bulge within its effective radius, in mag per
arcsec$^2$, Column 5 - bulge effective radius in kpc, Column 6 - the bulge
S\'{e}rsic index, Column 7 - disk central brightness in mag per arcsec$^2$,
Column 8 - disk scale length in kpc, Column 9 - integrated apparent magnitude
of the bar, Column 10 - S\'{e}rsic index of bar, Column 11 - Bar effective
radius in kpc, Column 12 - Bulge-to-Total luminosity ratio, Column 13 -
Bar-to-Total luminosity ratio.
\end{minipage}
\end{center}
\end{table*}

For the eight galaxies in the current sample, we have the photometric parameters
from \citet{Vaghmare2015}
obtained by modelling the mid-infrared surface brightness distribution. These parameters are reproduced in Table
\ref{tab:phot_params}. In Table \ref{tab:spec_params}, we have various spectral
parameters derived by fitting \textsc{starlight} models to the spectra from the central
regions of these galaxies. Also included in this table are $D_n(4000)$ index
\citep{Balogh1999}
values computed as the ratio of the
integrated fluxes from 4000 - 4100 \AA\ and 3850 - 3950 \AA\ bands. Using standard statistical techniques, we have studied 
correlations among photometric and spectroscopic 
parameters which we discuss in Section \ref{sec:corr_specphoto}.

Using the tools and techniques explained in the previous sections, we were able
to model the spectra from the central regions of the galaxies in our sample and
derive their various stellar population parameters. The central spectra and the best fit
\textsc{starlight} models are shown in Figure \ref{fig:spectra_all}. 
The star formation history as output by \textsc{starlight} is a
population vector which contains percentage contributions of SSPs of various
ages and metallicities, present in the base spectra, to the observed spectrum. To represent the star formation
history of the galaxy on a plot, we consider uniform bins of age in the log
space and use the population vector to determine the total percentage contribution by
all SSPs in a given age bin. So the plot is made with the logarithm of the age 
on the X axis and percentage of
light contributed by an SSP at that age along the Y-axis. The star formation
histories of the galaxies are shown in Figure \ref{fig:sfh_all} and the metallicity 
histograms in Figure
\ref{fig:metal_all}.

In Figure \ref{fig:grads_others1},
 we show the age and
metallicity gradients for the six unbarred galaxies and the two barred galaxies.
These plots are constructed as follows. For any given galaxy, we define the the
center using light weighted average row number for the brightness profile along
the slit. This is labelled as zero in all the plots. For each spectrum derived,
a light weighted centroid is assigned to it using the spatial profile in the
series of rows which were summed up to obtain the spectrum. \textsc{starlight} models
give us light weighted age and metallicity for this spectrum as well as
the errors on them. This way, one can obtain the gradients of the light weighted
age as well as the light weighted metallicity.

\subsection{Broad Trends}

In this section, we will discuss the broad trends found in this sample of eight
galaxies and reserve the next section for comments on individual objects. As
discussed earlier, the galaxies in this sample are a subset of the sample of
photometrically classified S0s hosting pseudobulges from \citet{Vaghmare2015}.
Pseudobulge hosting galaxies are in general expected to have a complex star
formation history viz. their spectrum cannot be described using that of a Simple Stellar Population.
This is because pseudobulges are believed to have been formed through slow and
secular processes with sustained star formation over a long period of time. An
SSP, on the other hand, describes a stellar population which has taken birth in
a single episodic burst.

The photometric studies on the current sample done by \citet{Vaghmare2015} 
reveal that their disks have a smaller scale length than their counterparts
among spiral galaxies. The authors argue that since the total luminosity of the
disk goes as $ L \sim I_d(0) r_d^2 $ the smaller scale length $r_d$ implies
that disks have a lower luminosity, assuming similar values of $I_d(0)$ which
they observationally confirm. They argue that this lower luminosity is due to
stripping of gas, a process often invoked to explain the fading of spiral arms
in spiral galaxies resulting in a morphological transformation to S0s. The
authors further argue in favor of gas stripping of spiral disks by comparing bulge luminosities.
Pseudobulges in S0s have a lower luminosity, which they say is due to the lower
availability of gas in S0s to drive bulge growth.

Based on these two known aspects of the current sample of S0s - that
they are pseudobulge hosts and they have undergone gas stripping, one
would expect two results from a detailed spectral study of such
galaxies. Firstly, we expect the spectrum of these galaxies to not be
well fitted by a single SSP spectrum.  Secondly, we expect no strong
indications  of recent star formation. Both these expectations can now be
tested using the spectral data obtained in the present study.

In Figure \ref{fig:spectra_all} we show the
final reduced spectra from the central regions of the eight galaxies.  It is
clear that the spectra of the non-barred galaxies (first six) indicate a population
dominated by young and recently formed stars while the spectra of barred
galaxies indicate an old and evolved stellar population likely to be well described
using an SSP. 
A spectrum of a typical elliptical galaxy, a class of galaxies
known to be consistent with having an SSP, are devoid of strong emission lines
and exhibit a strong break at the 4000 \AA\ region due to metal absorption
in the outer atmospheres of old stars. 
Such a break is seen in the spectrum of the two barred galaxies. But the break
is absent from the spectrum of five of six unbarred galaxies in Figure
\ref{fig:spectra_all}, indicating that the spectrum in their case is
dominated by contributions from a younger stellar population.

Further, one can see that most of the spectra show strong emission lines such as
$H\alpha$. This is counter to
the expectation that there should be no ongoing star formation in the current
sample of S0 galaxies, assuming that they are fully gas stripped and transformed
spirals.

In Figure \ref{fig:sfh_all}, we see the derived star
formation history of the central regions of these galaxies. One can again see that the
barred and the unbarred galaxies behaving differently. In case of
unbarred galaxies, we see peaks towards the lower end of the age axis indicating
that a large fraction of the light contributed to the observed spectrum comes
from the lower age population. We see the opposite result in the case of barred
galaxies which show higher peaks at the higher end of the age axis. An important
point to note regarding Figure \ref{fig:sfh_all}
concerns the lower limit on the age axis. The strong $H\alpha$ lines suggest
there must be a population of stars with $\log(\rm{age}) <7.0$ but the lowest
point on the age axis is 7.5. This is because the MILES spectral library does
not allow one to probe ages lower than this and not necessarily because there is
no population of age lower than the above limit. So, the lowest age bin should
be interpreted as containing percentage contribution from populations at that
age and lower.

Next, we explore the metallicity distributions in the unbarred and barred
galaxies, as shown in Figure
\ref{fig:metal_all} respectively. Again, it
is easy to spot the differences. In case of unbarred galaxies, there is a common
peak at 0.005 - 0.006 subsolar metallicity with little contribution from higher
metallicities. On the other hand, in case of barred galaxies, metallicities of
different ranges have comparable contributions to the stellar populations.

\begin{table*}
\caption{Best fit \textsc{starlight} parameters for the spectra from the central regions
of the eight galaxies in our sample.}
\label{tab:spec_params}
\begin{tabular}{l c c c c c c c c c c c c c}
\hline
Name & LWZ & MWZ & LWAge & MWAge & Av & Young & Inter & Old & $\textrm{Mass}_{\textrm{tot}}$ & $D_n(4000)$ \\
\hline
\hline
ESO079-007 & $ 0.007  $ & $ 0.009  $ & $ 8.275  \pm 0.039 $ & $ 9.706  \pm 0.053 $ & $ 0.488 \pm 0.056 $ & 0.768 & 0.0 & 0.232 & 8.429 & 1.076 \\
NGC1522    & $ 0.006  $ & $ 0.007  $ & $ 8.018  \pm 0.021 $ & $ 9.324  \pm 0.084 $ & $ 0.314 \pm 0.041 $ & 0.849 & 0.063 & 0.088 & 7.715 & 0.983 \\
NGC1510    & $ 0.006  $ & $ 0.008  $ & $ 8.029  \pm 0.021 $ & $ 9.356  \pm 0.094 $ & $ 0.282 \pm 0.035 $ & 0.754 & 0.194 & 0.052 & 7.917 & 1.002 \\
IC2085     & $ 0.009  $ & $ 0.013  $ & $ 8.953  \pm 0.025 $ & $ 9.961  \pm 0.03  $ & $ 0.892 \pm 0.019 $ & 0.429 & 0.03 & 0.541 & 8.125 & 1.329 \\
ESO085-030 & $ 0.008  $ & $ 0.01   $ & $ 8.627  \pm 0.019 $ & $ 9.82   \pm 0.048 $ & $ 0.582 \pm 0.022 $ & 0.597 & 0.0 & 0.403 & 8.285 & 1.201 \\
NGC7709    & $ 0.006  $ & $ 0.01   $ & $ 8.066  \pm 0.027 $ & $ 9.697  \pm 0.103 $ & $ 0.116 \pm 0.046 $ & 0.871 & 0.0 & 0.129 & 8.581 & 1.011 \\
NGC1533    & $ 0.02   $ & $ 0.02   $ & $ 10.176 \pm 0.01  $ & $ 10.224 \pm 0.003 $ & $ 0.042 \pm 0.011 $ & 0.018 & 0.0 & 0.982 & 9.065 & 1.731 \\
NGC5750    & $ 0.011  $ & $ 0.015  $ & $ 9.615  \pm 0.026 $ & $ 10.053 \pm 0.022 $ & $ 0.279 \pm 0.018 $ & 0.171 & 0.0 & 0.829 & 9.358 & 1.543 \\

 \hline
\end{tabular}

\textbf{Notes:} Column 1 - The common name of the galaxy;
Column 2 - Light Weighted Metallicity;
Column 3 - Mass Weighted Metallicity;
Column 4 - Log of Light Weighted Age [yr];
Column 5 - Log of Mass Weighted Age [yr];
Column 6 - Extinction Parameter;
Column 7 - Fraction of Young Stars (Age $< 10^8$ yr);
Column 8 - Fraction of Intermediate Stars ($10^8 < $ Age $ <10^9$ yr);
Column 9 - Fraction of Old Stars (Age $>10^9$ yr);
Column 10 - Log of Galaxy's Total Mass [$M_{\sun}$];
Column 11 - The $D_n(4000)$ index. The typical uncertainties in case of both
light and mass weighted metallicities are of the order of $10^{-4}$ and thus
have not been provided in the table above.

\end{table*}

Having summarised the broad trends in spectra, the star formation histories and
the metallicity distributions of the galaxies in the current sample, we now move
towards a more detailed description of individual galaxies.

\subsection{Discussion of Individual Objects}

\subsubsection{ESO079-007}

\begin{figure}
  \centering
  \includegraphics[width=0.8\columnwidth]{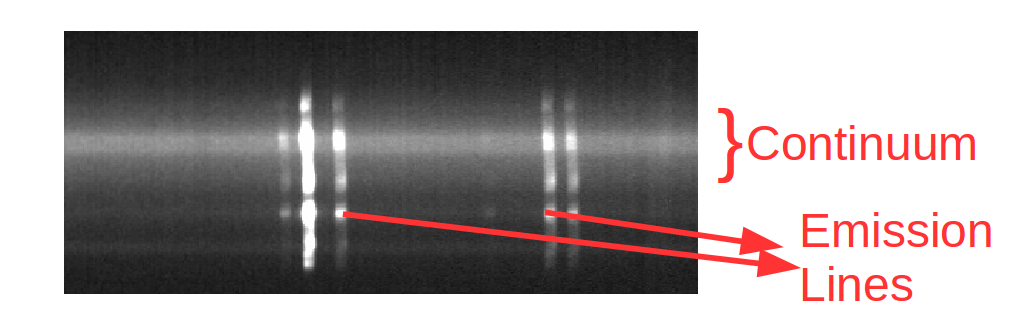}
  \caption{A narrow region of the 2-d spectrum of the galaxy ESO079-003, near
      the $H\alpha$, N[II] and S[II] emissions. The continuum has been marked along
  with the emission lines. As can be seen, the distribution of the emission
lines about the continuum is asymmetric.}
  \label{fig:eso079_asy}
\end{figure}

The NASA Extragalactic Database (NED), describes this S0 galaxy as being
peculiar. What sets this galaxy apart from conventional S0 galaxies is the
presence of asymmetric star forming regions i.e. the star forming regions are
more towards one side of the centre than the other. To illustrate this better,
we show in Figure \ref{fig:eso079_asy}, the 2-d spectrum near the wavelength
region dominated by $H\alpha$, N[II] and S[II] emission lines. The figure shows
the stellar continuum in the centre of the galaxy. It is easy to see that there
is a stronger star formation activity on one side of the galaxy centre than the
the opposite side. The overall metallicity of the galaxy is also sub-solar.

It is likely that the galaxy is accreting gas only from one direction
  in the disc which builds up the gas reservoir to induce disc
  instabilities and trigger asymmetric star formation \citep{Vulcani2018}. 
  Galaxy interactions in the form of fly-bys or
  recent accretion of smaller satellites could also account for the
  asymmetric distribution of star formation within the disc \citep{Mapelli2008}.

\subsubsection{NGC 1522} 

NED classifies this galaxy as a peculiar S0 galaxy. The spectrum of this galaxy
suggests a very young population and strong ongoing star formation dominating
the light budget at optical wavelengths. \textsc{starlight} models suggest that 95\%
of the flux is explained by a stellar population younger than $10^8$ yr. What
makes the object even more interesting is that it is highly metal poor
suggesting that the gas responsible for the ongoing star formation is not
enriched by any Population 2 stars.

Bulge-disk decomposition for this galaxy shows that the bulge component of the
stars accounts for 40\% of the total galaxy light and the rest by a faded disk. Yet again we
have a case of a galaxy whose star formation indicates availability of a large
amount of gas but the disk shows signs of having faded, assuming a spiral
progenitor with a brighter disk. 

\subsubsection{NGC 1533}
This  S0 galaxy is a part of the Dorado Group with a clearly discernable bar structure. It exhibits
a smooth appearance and one look at the spectrum of the galaxy reveals that it
is dominated by a very old population. An age gradient plot reveals that indeed,
95\% of the galaxy light is dominated by old stars. From 3.6 micron imaging, we
know the bulge effective radius to be $\sim 4$''. It has been pointed out in the
literature \citep{Gadotti2009} that bulge light dominates the disk light for
atleast twice the bulge effective radius. So, it reasonable to expect that to
obtain a disk dominated spectrum sufficiently free from any bulge light
contamination, one would have to go twice as far (16'').
This means we do not have much
signal from the disk dominated region to be able to compare the populations of
bulges and disks. The metallicity is also near solar which is not too surprising
given other features of this galaxy.

This galaxy is clearly quite different from the first two examples in our
sample - devoid of star formation and dominated by older stars. It is
also different in another major way - it is a barred galaxy. Its formation
history is consistent with an old coeaval population but with a small component
of recently formed stars. These stars could have formed through gas infall
induced by bars. 

\subsubsection{NGC 1510}
This galaxy is known to be a part of a pair with the other galaxy
  being NGC 1512. According to \citet{Muerer2006}, this is a pair of
  galaxies exhibiting starbursts and a common HI envelope. The
  presence of an intense star burst is confirmed by our spectrum which
  shows extremely strong $H\alpha$ emission. The age variation as a
  function of distance from the centre reveals an asymmetric
  feature. This means that there are more old stars on one side of the
  galaxy. This population of old stars on one side is also metal rich
  as can be seen in Figure \ref{fig:grads_others1}.

These features seem to be a result of the tidal effects of its
  partner galaxy NGC 1512. The tidal features are clearly visible in
  NGC 1510 from archival {\it GALEX} FUV and NUV images.  The tidal interactions
  have likely led to the asymmetric formation of star burst regions in
  this galaxy. Again, the B/T ratio in this galaxy is quite high at
  0.4.

\subsubsection{IC 2085}
This galaxy is part of NGC~1566 group of galaxies which is a part
  of the Dorado Group. In comparison to other galaxies in this sample, the spectrum 
obtained for this galaxy is of a relatively lower SNR. Even without 
performing any modelling, one can see that this galaxy contains 
both a very old and a very young stellar population in the centre. The 
presence of strong emission lines such as $H\alpha$ support ongoing star 
formation as well as recently formed stars while the 4000 \AA\   break caused by
several absorption features below 4000 \AA\ 
is indicative of an older population.   Figure 9  shows an asymmetric 
age gradient around the centre however the stellar metallicity
distribution  is symmetric. This could be due to the bursts
of star formation in the presence of asymmetric dust features. The
optical image from Digital Sky Survey (DSS) confirms the presence of
disturbed dust lanes.

The best-fit model from \textsc{starlight} reveals that 50\% of the light is accounted for by a
young population  with age $<10^8$ yr and the rest by very old stars. 
Such a mixture of very young and very old stars hints at a composite
bulge system.

\subsubsection{ESO085-030}
This galaxy shares several features with ESO079-007 except for
  the star forming regions which are not distributed asymmetrically.
  About 40\% of the light in the centre of this galaxy is being
  accounted for by old stars and this fraction falls to 15\% as one
  moves outwards. The B/T ratio as determined from the 3.6-micron
  imaging is 0.6 with the disk scale length smaller than the bulge
  effective radius and the Sersic index is highest in the sample. This
  means that at least in the centre, the bulge which is surrounded by the
  small disk dominates significantly and thus the trend being observed
  is due to the bulge. This indicates that pseudobulges grow in a
  manner similar to disks i.e. inside-out.

That the pseudobulge can have a B/T ratio of 0.6 needs further explanation here.
Based on prior studies carried out on pseudobulges, one generally does not expect
such a high value of B/T. These studies have largely focussed on late-type
galaxies which are known to be dominated by disks. However, it is possible that
in the case of an S0 galaxy, the disk has lowered in luminosity due to a
morphology transformation through gas stripping. This would cause an increase in
the B/T ratio of these objects.

\subsubsection{NGC 5750}
This is an interesting object which exhibits a ring like structure. It is
classified as an S0/a galaxy which designates it as a transition class object,
i.e. an object which might be speculated to be transforming between S0 and Sa
morphology. Sandage, in \emph{The Carnegie Atlas of Galaxies} speculates that
the ring might actually be made of two very tightly bound spiral arms. 

The galaxy is barred and in terms of spectral properties, is similar to NGC
1533, comprising of an old population with near solar metallicity. It is
also the second barred galaxy in our sample which has been classified as a
pseudobulge based on photometric properties, but it shows characteristics that
are atypical and not expected of pseudobulges.

\subsubsection{NGC 7709}
All conclusions about the NGC 1522 apply to this galaxy as is - there is
evidence of strong ongoing star formation, very little flux contribution from
the old stars and an overall sub-solar metallicity.

\subsection{Correlations between Photometric and Spectroscopic Parameters}
\label{sec:corr_specphoto}

\begin{figure}
  \centering
  \includegraphics[width=0.8\columnwidth]{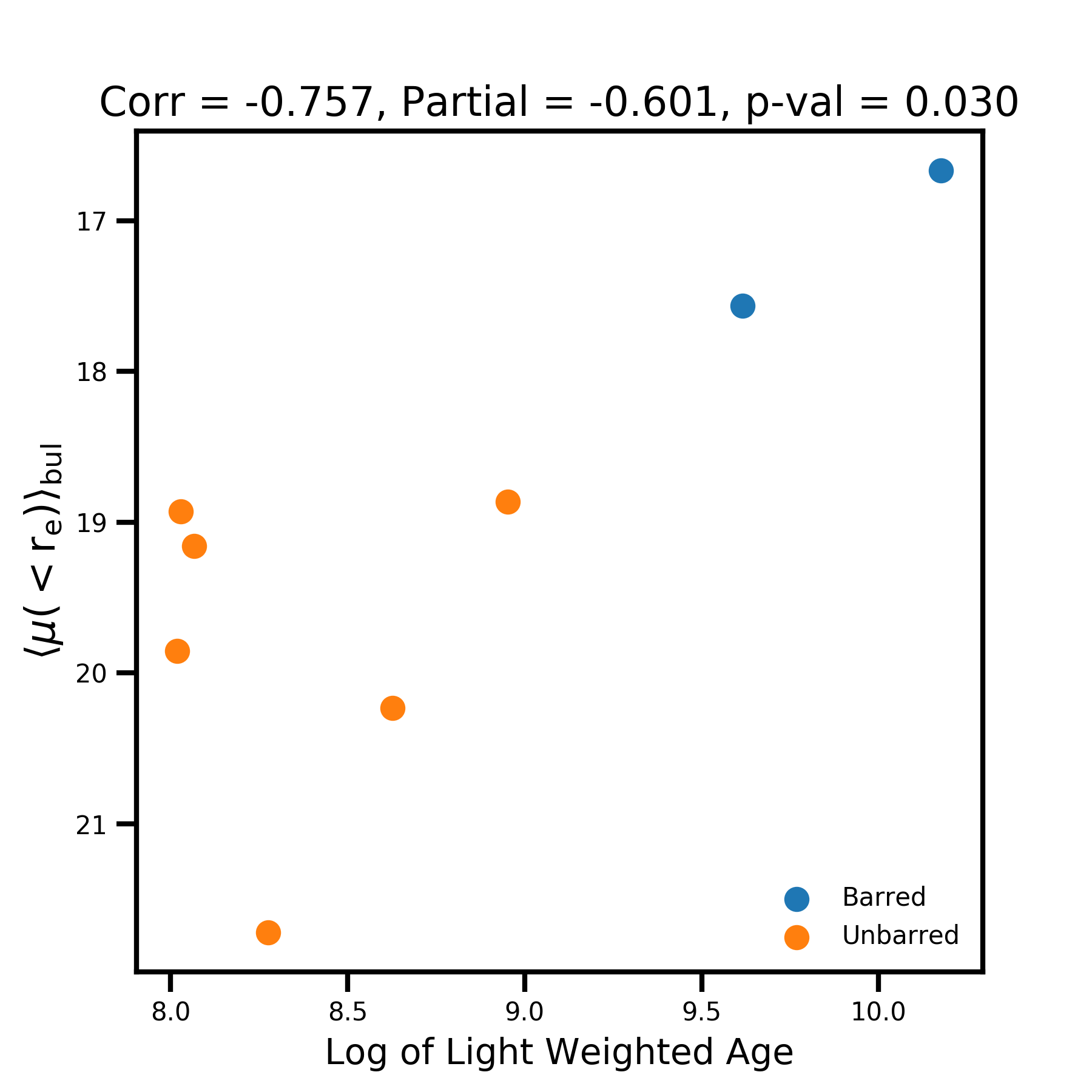}
  \includegraphics[width=0.8\columnwidth]{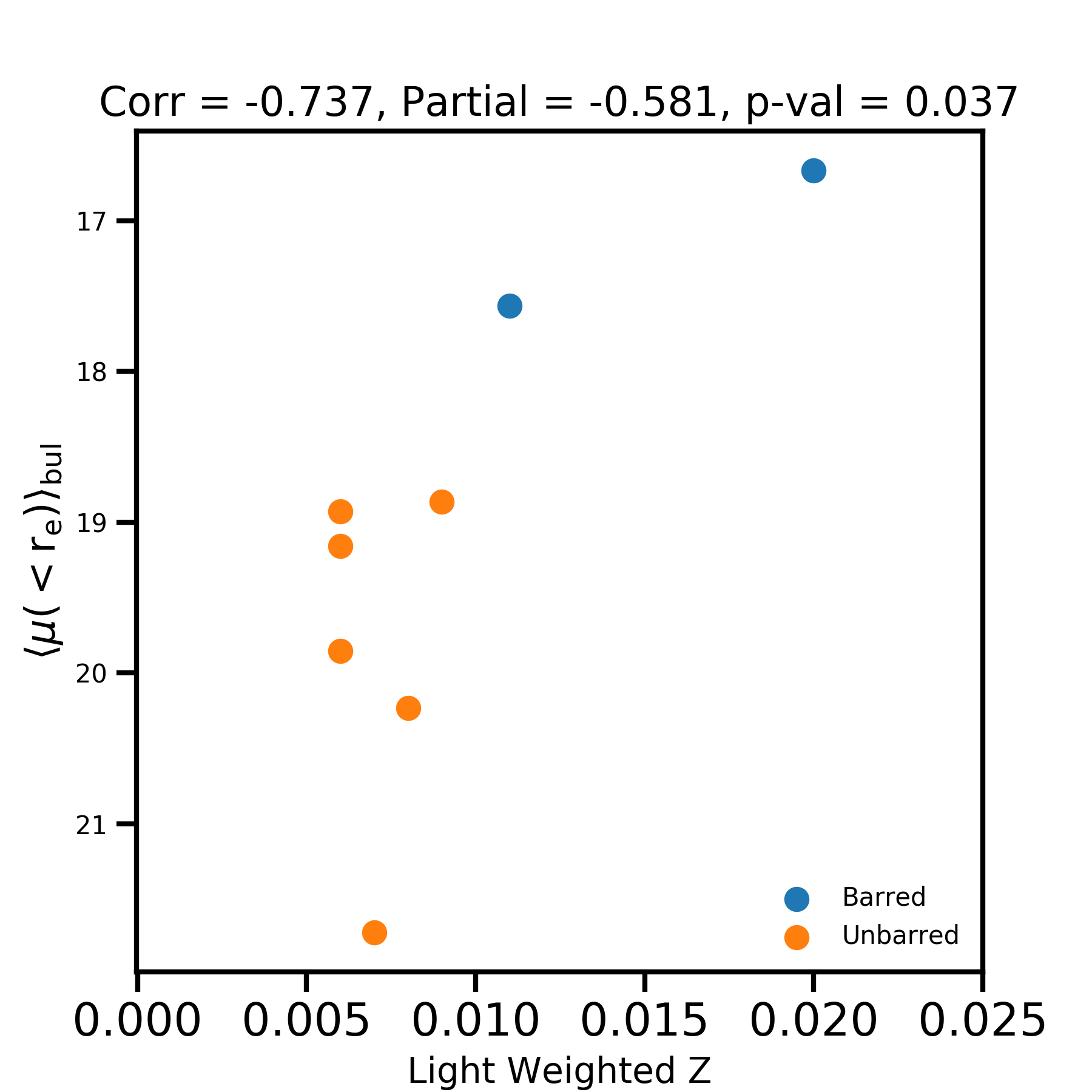}
  \caption{A plot of bulge average surface brightness within effective radius as
  a function of light weighted age and metallicity.}
  \label{fig:bul_trend}
\end{figure}

In Figure \ref{fig:bul_trend}, are two plots showing the variation of bulge
average surface brightness within effective radius against light weighted age and
metallicity respectively. We see a clear trend that average surface brightness
increases with increasing age and metallicity.
All other correlations show significant scatter and thus have not been
discussed here. We would like to point out that the present study cannot conclusively
determine whether the scatter is because of low number statistics or indicates
an absence of a real correlation altogether. Further investigation with a larger sample
is warranted.

The Spearman rank correlation coefficient between average surface
brightness of the bulge within its effective radius and the logarithm of the
light weighted age obtained from the \textsc{starlight} model, is $-0.757$ with a
significance of $ > 95$ \%. A similar correlation is seen with metallicity; the rank
correlation for this is found to be  $-0.737$  with similar significance.
These correlations point at a relation between age and metallicity of
the population and the growth / build up of bulges.

These correlations could also be driven by the galaxy mass. For example, if the
age or metallicity are correlated with the mass of the galaxy
\citep{Gallazzi2005}, then one should
expect to see the correlation we find in the upper panel of Figure
\ref{fig:bul_trend}. A simple way to
test the possibility is to evaluate the Spearman partial rank correlation
between the average surface brightness and age, with the effect of total mass of
the galaxy factored out. The masses of our galaxies were determined by a
prescription given in \citet{Cook2014} from the M/L ratios at 3.6
$\mu$m. We find that the partial correlation coefficient is
$-0.601$, which is not significantly different from the Spearman rank
correlation coefficient between the two variables, which is $-0.757$. Similarly,
the partial correlation coefficient between the average surface brightness and
metallicity is $-0.581$, which is comparable to the correlation coefficient if
$-0.737$. This seems to suggest that the correlations are not completely driven
by the total mass of the galaxy. Note that these correlations hold even if we
use the corresponding mass weighted age and metallicity, albeit with a very minor 
change in the correlation strengths.

Again, we would like to point out that these results are not statistically
robust owing to the small number of data points. But the presence of such
correlations suggests the need for a study which aims at obtaining both
photometric and spectroscopic parameters for a larger sample of galaxies.

\section{Discussion}
\label{sec:discussion}

We have seen above that the barred galaxies in our sample have largely similar
stellar population properties and so is the case with the unbarred galaxies
taken as a group but there are important differences between these
groups.

\subsection{The unbarred group}
\label{subsec:unbarred_group}

Recall that the parent sample of this study is the sample of 25 pseudobulge
hosting S0 galaxies, as described in \citet{Vaghmare2013}. The authors
classified pseudobulges using a very conservative criterion. Typically, studies
aimed at bulge classification adopt a very simple criterion where all bulges
with S\'{e}rsic index $n<2$ are classified as pseudobulges and others as
classical. But as \citet{Vaghmare2013} argue, the parameter $n$ is not easy to
constrain and thus there is scope for misclassification. \citet{Gadotti2009}
proposed an alternate criterion based on the Kormendy diagram and argued on the
merits of such an approach. \citet{Vaghmare2013} adopted a combination of the
two criteria to ensure that a secure classification is obtained. The chances of
misclassifying a galaxy as a pseudobulge host is thus quite low. 

\citet{Vaghmare2015} suggested that the pseudobulge hosting S0s have a
lower disk luminosity with respect to their spiral counterparts, which
the authors interpret as a disk that has faded while undergoing a
transformation in morphology from spiral to S0. The authors
argue that this transformation involves stripping of gas which is
needed for sustaining active star formation. The signs of active star
formation in these galaxies, as indicated by their spectra is in
contradiction with this conclusion. Further to confirm if the gas
stripping process is a viable scenario,  we investigated the
environment of these unbarred galaxies. We found that excluding  
IC\,2085, the galaxies in the unbarred group are in an isolated
environment. 

One possible resolution is that these isolated galaxies are not
 gas stripped spirals. In this case, we have to conclude that these
 disks are inherently less luminous and somehow such disks have an 
affinity to be a part of pseudobulge hosting S0 galaxies. Another 
explanation is that these objects are gas stripped spirals but have 
had their gas replenished through numerous gas rich minor mergers
which in turn led to complex star formation \citep{Penoyre2017}.

\subsection{The barred group}
\label{subsec:barred_group}

An interesting commonality in the two galaxies in this group, apart from the
stellar population, is that these galaxies are barred. With the usual caveat
of lack of robustness due to small numbers, this common observation allows for
some speculation on why, despite such a conservative classification criterion, the
bulges of these galaxies are found not to be consistent with pseudobulges.
Recall that the photometric parameters of the bulges as derived by performing
2-d bulge-disk decomposition, have been used for classifying them. One may thus
argue that the presence of a bar has a systematic effect on the derived bulge
properties \citep{Laurikainen2005, Gadotti2009}. But \citet{Vaghmare2013}
recognize this problem and take care to fit a bar component as well. Thus
the unaccounting of a bar component cannot explain this
observation.

Another explanation for the old stellar population in these galaxies is as
follows. The presence of the bar led to an amplified funnelling of gas
towards the centre of the galaxies in the past. This allowed the pseudobulge
to form much earlier than the rest of the galaxies where there was no bar
component to aid the quick formation of a pseudobulge. As a result the
pseudobulges in both these galaxies contain a very old population as can be seen
in the Figure \ref{fig:sfh_all}. In the case of NGC 1533, there does not seem to
have been any gas available to lead to recent star formation while in the
case of NGC 5750, there is some evidence of recent star formation.
A confirmation of whether the bar action can explain the older age could be
obtained from a  high spatial resolution rotation curve for these galaxies. The
relative domination of a random component of stellar velocities vs systematic
rotation may be able to shed light on whether there is funnelling of gas taking
place. However, such a study is not possible using the present data.

\section{Stellar Populations in Pseudobulge Hosting S0s - Barred vs Unbarred}
\label{sec:bar_dn4000}

In section \ref{subsec:barred_group}, we saw that the only pseudobulge hosting S0 galaxies
which exhibited a significant departure from the overall nature of
star formation history observed among other galaxies were
barred. However, with such low numbers, it is not appropriate to firmly conclude
that the nature of pseudobulges in S0 galaxies is a function of the
presence or absence of a bar. But it is reasonable to expect a
difference since we know that the bar does affect the nature of
stellar orbits and gas dynamics in a galaxy. Further, a bar is known
to be a driver of growth in bulges in the secular evolution scenario.

We thus decided to perform a quick analysis to see if such a 
difference between pseudobulges in barred S0s and unbarred S0s, is indeed
real. \citet{Mishra2017} have presented evidence for the bimodal
stellar age distribution of pseudo-bulges of unbarred S0 galaxies as
probed by the $D_n(4000)$ index. This index is a reliable indicator of
the mean age of galaxy stellar population which is quantified using
the strength of the 4000 \AA{} spectral break arising due to the
accumulation of absorption lines of metals in the atmosphere of old,
low-mass stars in galaxies. The light-weighted mean stellar ages of
$\sim$2 Gyr corresponds to $D_n(4000)$ index of 1.5  \citep{Kauffmann2003}
and using this \citet{Mishra2017} have divided the sample of
unbarred pseudo-bulges in S0 galaxies with a young (having $D_n(4000) <
1.5$)  and old (having $D_n(4000) \ge 1.5$) stellar populations. In
their study, Mishra et al. (2017) did not use S0 galaxies that host a
bar. In this work, we use the sample and data described in \citet{Mishra2017} 
but for pseudo-bulges in S0 galaxies that host a bar.  

\begin{figure}
  \centering
  \includegraphics[width=0.8\columnwidth]{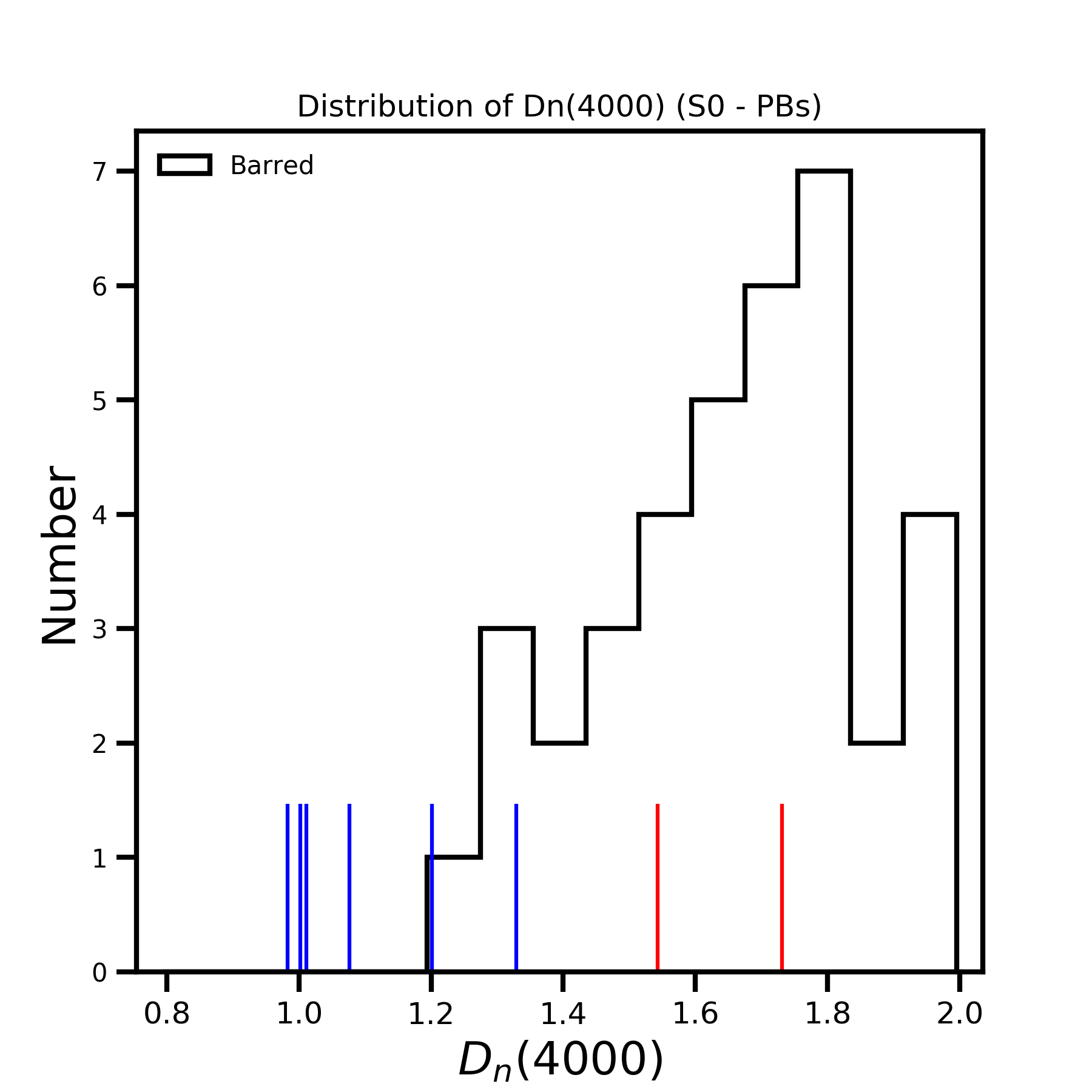}
  \caption{A plot showing the distribution of $D_n(4000)$ index for
  barred pseudobulge hosting S0 galaxies from the sample used in
  \citet{Mishra2017}. The blue vertical lines represent 
  the $D_n(4000)$ index for unbarred galaxies in our sample while the red
  vertical lines represent the barred galaxies.}
  \label{fig:barwise_dn4000}
\end{figure}

In Figure \ref{fig:barwise_dn4000}, we show the distribution of the $D_n(4000)$ index for all
the barred S0s with pseudobulge from the sample in
\citet{Mishra2017}. This figure reveals
that the barred galaxies preferentially have  $D_n(4000) \ge 1.5$ ( older stellar
populations), which is consistent with our current study. The figure also shows the $D_n(4000)$ index
for the eight galaxies in our sample. Our current spectroscopic analysis seems to suggest that the bimodality
presented in \citet{Mishra2017} could be caused by the presence of a
bar. This, in turn, suggests that pseudobulges, when found in barred
S0s exhibit a dominant older stellar population
while an unbarred pseudobulge hosting S0 can exhibit significant contribution from 
both young and old populations. Our current spectroscopic sample,
however, does not have pseudobulge hosting S0s which have a
very old population and do not contain a bar. Commenting on the exact nature of
such differences in S0s with pseudobulge however, is rather difficult.

\section{Summary}
\label{sec:summary}
In this paper, we have studied eight S0 galaxies established by
\citet{Vaghmare2015} to host pseudobulges, using long-slit spectra obtained from
the SALT-RSS. We have modelled the spectra using \textsc{starlight} to derive detailed
star formation histories of these galaxies as well as age and metallicity
gradients. The present study focusses primarily on the spectra of the central
regions of the galaxy and, using the spectral information, confronts the conclusions drawn by \citet{Vaghmare2015}.

The most important conclusion is that these eight galaxies are not all similar
in terms of their star formation histories. This
means that objects grouped in one class using photometric criteria
(the class here being pseudobulge hosting S0s) can exhibit diverse properties.
The present study brings out the need for further studies that are based on
spectral analysis to truly constrain the formation mechanisms at work in
the formation of these objects.

We discovered that six of these eight galaxies, which are unbarred, exhibit a
complex star formation history with active star formation. We attempt
reconciling the observed ongoing star formation with the conclusion reached by
\citet{Vaghmare2015} that these are gas stripped spirals. It has been suggested
by \citet{Poggianti2017} that the process of ram pressure stripping in galaxies
can cause initial fuelling of gas inwards, which could lead to star formation in
the central regions. Once the gas is stripped, the disk can fade, but residual
gas in the central region could be responsible for the observed star formation.
This would make fading due to stripping consistent with our observations. But
since most of the galaxies are in an isolated environment the question would
remain as to what caused the stripping.

In the other two galaxies, which are barred, we find a population similar to
those expected in elliptical galaxies. We have detected a trace of recent star
formation in one of these galaxies which likely indicates that the bar is
funnelling / has funnelled gas towards the center leading to a formation of a
younger group of stars.
We have also used a
sample of SDSS galaxies to show that the preferential occurrence of older populations
in barred S0 galaxies, suggested by the spectral data, is likely a real
phenomenon. The study of this SDSS sample also suggests that there are
unbarred pseudobulge hosting S0s with very old populations \citep{Mishra2017}.
However, no such object
has been found in the present study. 

Using the data obtained from the SALT-RSS, it is possible to perform a very
detailed analysis of the star formation history of the off-center regions of the
galaxy. This can lead to new insights into how galaxies formed. 
We defer such a detailed study to a future paper.

\section*{ACKNOWLEDGEMENTS}
We thank the anonymous referee for insightful comments that
have improved both the content and presentation of this paper. 
All of the observations reported in this paper were obtained with the
Southern African Large Telescope (SALT) under program 2014-1-IUCAA-RSA-OTH-001 (PI:
A.K. Kembhavi). KV would like to acknowledge Alexei Kniazev, Steve Crawford and R
Srianand for useful discussions and guidance. KV also acknowledges 
the Council of Scientific and Industrial Research (CSIR), India for
financial assistance and SAAO \& SALT for travel and hospitality
support. Part of this work for KV was possible due to a grant from the National
Knowledge Network (NKN), India. SB acknowledges SAAO where part of this work 
has been done. PV acknowledges support from the SA National Research Foundation.
YW thanks IUCAA for hosting him on his sabbatical where a part of this work was done. We acknowledge support from a South African National 
Research Foundation grant (PID-93727) and from a bilateral grant under 
the Indo-South Africa Science and Technology Cooperation (PID-102296) 
funded by the Department of Science and Technology (DST) of the Indian 
and South African governments. PyRAF is a product of the Space Telescope Science
Institute, which is operated by AURA for NASA and was extensively utilised for
the spectral data reduction pipelines.



\bibliographystyle{mnras}

\section*{Appendix A}

As mentioned in Section \ref{sec:specmod}, we construct the base spectra starting
from a collection of 150 SSP spectra from the MILES library. These spectra are
then input to the algorithm proposed by \citet{Richards2009} and a new set of 45
base spectra are obtained. Table \ref{tab:base} shows the ages and the
metallciites of the final base spectra used with \textsc{starlight}.

\begin{table}
    \caption{Ages and metallicities of the base spectra used in the present
    study.}
    \label{tab:base}
    \begin{tabular}{l c c}
    \hline
    S.No. & Age (yr) & Metallicity \\
    \hline
1 & $ 2.6396 \times 10^{ 8 }$ & $ 6.8388 \times 10^{ -3 }$ \\ 
2 & $ 1.3111 \times 10^{ 10 }$ & $ 1.3994 \times 10^{ -2 }$ \\ 
3 & $ 9.7615 \times 10^{ 9 }$ & $ 1.0982 \times 10^{ -2 }$ \\ 
4 & $ 2.6210 \times 10^{ 8 }$ & $ 1.0380 \times 10^{ -2 }$ \\ 
5 & $ 8.9564 \times 10^{ 8 }$ & $ 1.0630 \times 10^{ -2 }$ \\ 
6 & $ 5.6230 \times 10^{ 8 }$ & $ 1.0638 \times 10^{ -2 }$ \\ 
7 & $ 4.8752 \times 10^{ 8 }$ & $ 8.9622 \times 10^{ -3 }$ \\ 
8 & $ 8.9130 \times 10^{ 8 }$ & $ 4.0000 \times 10^{ -3 }$ \\ 
9 & $ 7.9791 \times 10^{ 7 }$ & $ 5.9980 \times 10^{ -3 }$ \\ 
10 & $ 1.8109 \times 10^{ 9 }$ & $ 1.0634 \times 10^{ -2 }$ \\ 
11 & $ 6.1258 \times 10^{ 9 }$ & $ 6.6698 \times 10^{ -3 }$ \\ 
12 & $ 1.2589 \times 10^{ 10 }$ & $ 2.0000 \times 10^{ -2 }$ \\ 
13 & $ 7.3106 \times 10^{ 9 }$ & $ 1.5997 \times 10^{ -2 }$ \\ 
14 & $ 5.4542 \times 10^{ 9 }$ & $ 8.9952 \times 10^{ -3 }$ \\ 
15 & $ 9.3783 \times 10^{ 7 }$ & $ 1.0016 \times 10^{ -2 }$ \\ 
16 & $ 7.0790 \times 10^{ 8 }$ & $ 2.0000 \times 10^{ -2 }$ \\ 
17 & $ 7.3706 \times 10^{ 9 }$ & $ 9.9992 \times 10^{ -3 }$ \\ 
18 & $ 7.5109 \times 10^{ 8 }$ & $ 6.0003 \times 10^{ -3 }$ \\ 
19 & $ 4.2302 \times 10^{ 9 }$ & $ 8.8058 \times 10^{ -3 }$ \\ 
20 & $ 3.9137 \times 10^{ 9 }$ & $ 6.0026 \times 10^{ -3 }$ \\ 
21 & $ 7.9430 \times 10^{ 8 }$ & $ 8.0000 \times 10^{ -3 }$ \\ 
22 & $ 2.9023 \times 10^{ 9 }$ & $ 9.3242 \times 10^{ -3 }$ \\ 
23 & $ 2.1840 \times 10^{ 8 }$ & $ 5.3372 \times 10^{ -3 }$ \\ 
24 & $ 1.0052 \times 10^{ 9 }$ & $ 1.0613 \times 10^{ -2 }$ \\ 
25 & $ 1.4125 \times 10^{ 10 }$ & $ 2.0000 \times 10^{ -2 }$ \\ 
26 & $ 1.4844 \times 10^{ 9 }$ & $ 1.0627 \times 10^{ -2 }$ \\ 
27 & $ 7.4950 \times 10^{ 9 }$ & $ 9.0092 \times 10^{ -3 }$ \\ 
28 & $ 3.3522 \times 10^{ 9 }$ & $ 1.2004 \times 10^{ -2 }$ \\ 
29 & $ 1.1220 \times 10^{ 10 }$ & $ 2.0000 \times 10^{ -2 }$ \\ 
30 & $ 6.5028 \times 10^{ 8 }$ & $ 8.9704 \times 10^{ -3 }$ \\ 
31 & $ 1.3282 \times 10^{ 8 }$ & $ 1.8674 \times 10^{ -2 }$ \\ 
32 & $ 1.7903 \times 10^{ 9 }$ & $ 1.3993 \times 10^{ -2 }$ \\ 
33 & $ 3.9982 \times 10^{ 8 }$ & $ 5.3293 \times 10^{ -3 }$ \\ 
34 & $ 1.7893 \times 10^{ 9 }$ & $ 6.0078 \times 10^{ -3 }$ \\ 
35 & $ 1.6733 \times 10^{ 10 }$ & $ 2.0000 \times 10^{ -2 }$ \\ 
36 & $ 3.2165 \times 10^{ 8 }$ & $ 8.9766 \times 10^{ -3 }$ \\ 
37 & $ 1.0998 \times 10^{ 10 }$ & $ 1.4071 \times 10^{ -2 }$ \\ 
38 & $ 2.3371 \times 10^{ 9 }$ & $ 8.9891 \times 10^{ -3 }$ \\ 
39 & $ 3.2452 \times 10^{ 9 }$ & $ 6.0036 \times 10^{ -3 }$ \\ 
40 & $ 6.3100 \times 10^{ 7 }$ & $ 5.9987 \times 10^{ -3 }$ \\ 
41 & $ 1.1220 \times 10^{ 9 }$ & $ 2.0000 \times 10^{ -2 }$ \\ 
42 & $ 1.4188 \times 10^{ 9 }$ & $ 5.3253 \times 10^{ -3 }$ \\ 
43 & $ 1.0286 \times 10^{ 10 }$ & $ 1.2000 \times 10^{ -2 }$ \\ 
44 & $ 2.0716 \times 10^{ 8 }$ & $ 1.3105 \times 10^{ -2 }$ \\ 
45 & $ 1.1273 \times 10^{ 9 }$ & $ 1.0638 \times 10^{ -2 }$ \\ 

\hline       
    \end{tabular}

    Notes: The base spectra whose ages and metallicities are given above have
    been constructed by inputting 150 initial base spectra from the MILES
    library to the \textsc{MATLAB} code which implements the algorithm by
    \citet{Richards2009}.
\end{table}



\label{lastpage}
\end{document}